\documentclass[footinbib,twocolumn,showpacs,amsmath,amstex,amssymb,mathfonts,superscriptaddress,prx]{revtex4}
\usepackage{graphicx}
\usepackage{color}
\usepackage{bm}

\usepackage{graphicx}
\usepackage{color}
\usepackage{bm}
\usepackage{amsmath}
\usepackage{amssymb}
\usepackage{amsthm}
\usepackage{amsfonts}
\usepackage{multirow}
\usepackage{makecell}
\usepackage{float}
\usepackage{comment}
\usepackage{enumitem}
\usepackage{verbatim}
\usepackage{mathtools}

\usepackage{bbm}

\begin{document}

\title{The composite fermion theory revisited: a microscopic derivation without Landau level projection}
\author{Bo Yang} 
\affiliation{Division of Physics and Applied Physics, Nanyang Technological University, Singapore 637371.}
\affiliation{Institute of High Performance Computing, A*STAR, Singapore, 138632.}

\date{\today}
\begin{abstract}
The composite fermion (CF) theory gives both a phenomenological description for many fractional quantum Hall (FQH) states, as well as a microscopic construction for large scale numerical calculation of these topological phases. The fundamental postulate of mapping FQH states of electrons to integer quantum Hall (IQH) states of CFs, however, was not formally established. The Landau level (LL) projection needed for the microscopic calculations is in some sense uncontrolled and unpredictable. We rigorously derive the unitary relationship between electrons and the CFs, showing the latter naturally emerge from special interactions within a single LL, without resorting to any projection by hand. In this framework, all FQH states topologically equivalent to those described by the conventional CF theory (e.g. the Jain series) have exact model Hamiltonians that can be explicitly derived, and we can easily generalise to FQH states from interacting CFs. Our derivations reveal fundamental connections between the CF theory and the pseudopotential/Jack polynomial constructions, and argue that all Abelian CF states are physically equivalent to the IQH states, while a plethora of non-Abelian CF states can be systematically constructed and classified. We also discuss about implications to experiments and effective field theory descriptions based on the descriptions with CFs as elementary particles.  
\end{abstract}

\maketitle 

\section{Introduction}

Recently it has been shown that anyons (including fermions) in a single Landau level (LL) of the fractional quantum Hall (FQH) effect can be bosonized\cite{PhysRevLett.127.126406}. This implies that any FQH states, including the ground states and the quasiholes, can be understood as quantum fluids of bosonic degrees of freedom. This is fundamentally due to the conformal invariance of the Hilbert spaces in a single LL, and the bulk-edge correspondence within these Hilbert spaces that can connect bulk states to the edge excitations described by the chiral Luttinger liquid theory\cite{PhysRevB.79.245304,yang2021gaffnian}. Thus statistical transmutation can be naturally performed between any two types of anyons, and we can in principle write down effective field theories for the same topological phase with either fermionic or bosonic degrees of freedom.

The ability of bosonization of any FQH quantum fluids also implies we can fermionize these quantum fluids. The fundamental degrees of freedom of all systems we are interested in here are electrons. Thus to identify emergent bosons or fermions from the many-body wavefunctions of electrons, we need to make sure these emergent particles are countable, and they have an unambiguous set of quantum numbers just like the electrons. Without loss of generality, in this work we will focus on the lowest Landau level (LLL) on the spherical geometry\cite{PhysRevLett.51.605}. For electrons, if we fix the strength of the magnetic monopole at the center of the sphere to be $2\bm S$, then each electron can be understood as a spinor with total ``spin" $\bm S$, which is equivalent to its total angular momentum on the sphere. Throughout this paper we will be dealing with \emph{spinless fermions}, so we use the term ``spin" to denote the total angular momentum of the particle to emphasise its spinor structure on the spherical geometry. On this two-dimensional manifold the single particle state has two quantum numbers, and it is an eigenstate of the total angular momentum operator $\hat L^2$, and the z-component of the angular momentum operator $\hat L_z$. It is important to note, however, that no matter how many electrons are added to the sphere, each electron has the same spin (the eigenvalue of $\hat L^2$) $\bm S$ in the LLL, independent of the number of electrons present. Each electron can thus be indexed with a quantum state $|m,\bm S\rangle$, with $m$ running from $-\bm S$ to $\bm S$. One should note the total number of orbitals in the LLL is given by $2\bm S+1$. The fully filled LLL thus contains $2\bm S+1$ electrons and is represented by the product state $|-\bm S,\bm S;-\bm S+1,\bm S;\cdots,\bm S-1,\bm S;\bm S,\bm S\rangle$.

The description above can be applied to any single LL; for the $n^{\text{th}}$ LL we just need to replace $\bm S$ with $\bm S+n$. In the bosonization scheme\cite{PhysRevLett.127.126406}, the magnetic fluxes inserted to the quantum fluid on the sphere can be treated as bosons if we fix the number of electrons $N_e$ instead (in contrast to the monopole strength being fixed for electrons). In this way, no matter how many fluxes are inserted, each flux can be treated as a boson, or a spinor on the sphere with total spin $S=N_e/2$. Each boson is thus a quantum state $|m,S\rangle$ with $m$ running from $-S$ to $S$, and the bosonic nature is revealed when we look at the allowed quantum numbers of the multi-boson states\cite{PhysRevLett.127.126406}. For example with two bosons, the allowed total spins are given by $2S-k$ with $k$ being a non-negative even integer. In contrast for two electrons, it is fermionic because the allowed total spins are $2\bm S-k$ with $k$ being a positive odd integer.

The ability to map an anyonic or fermionic basis to a bosonic basis immediately implies we can map an anyonic basis in the FQH fluids to a fermionic basis. Indeed, this is the underlying concept of the composite fermion theory\cite{jain1989composite,jain2007composite}, which reinterpret many experimentally observed FQH states as integer quantum Hall (IQH) states of a new type of fermionic degrees of freedom: the composite fermions (CF). Phenomenologically, each CF is a bound state of one electron with an even number of vortices. While the interaction between electrons are strong, the effective interaction between CFs are conjectured to be weaker, which is supported by finite size numerical calculations\cite{PhysRevB.94.165303,PhysRevB.96.245142}. The CF theory is very successful in explaining many experimental observations. It also leads to effective field theory descriptions of the FQH phases\cite{PhysRevB.47.7312,Lopez_1998}, by using CFs as the elementary Fock space degrees of freedom, together with a number of assumptions built in the original CF theory.

The conventional CF theory is most useful with the Coulomb interaction in the LLL, which is one of the most experimentally relevant regimes. The famous Jain series of the abelian FQH states observed in the experiments are well described by the IQH of CFs, presumably because with the LLL Coulomb interaction, the interaction between the CFs are weak compared to the emergent incompressible gaps at those FQH states. In higher LLs, the effective Coulomb interaction becomes long ranged, so that the short range part of the interaction is no longer dominant. There, the CFs become strongly interacting just like electrons, and thus the CF description of the FQH states becomes cumbersome. This is also true for many theoretically predicted non-Abelian FQH states, which are some of the most interesting aspects of the FQH physics that may also lead to universal topological quantum computing\cite{moore1991nonabelions,PhysRevB.75.075318,anyonrmp}. Model Hamiltonians of these non-Abelian states induce strong interaction between CFs, and more complicated schemes involving partons are developed in understanding these exotic topological phases\cite{PhysRevB.40.8079}. For these systems and other more exotic FQH phases, it is increasingly more difficult to justify the FQH to IQH correspondence via numerical analysis, and there are few rigorous understandings on conditions under which such correspondence is valid. 


\textit{The Landau level projection--}
From a technical perspective, the CF theory can readily generate many-body wavefunctions by attaching vortices (e.g by the multiplication of the factors $\prod_{i<j}\left(z_i-z_j\right)^n$) to an IQH wavefunction. This is followed by a very specific operation in the CF theory, which is the projection into the LLL, or the so-called LLL projection. While the original IQH states are orthonormal and form a complete basis, the resulting projected FQH many-body wavefunctions are generally not. This is because the LLL projection is \emph{not a unitary transformation} from the electron basis to the CF basis. The central assumption of the CF theory is a one-to-one correspondence between the projected and unprojected CF wavefunctions, at least for the states that are physically important. Such an assumption entirely relies on the numerical checking of finite system sizes based on the short range (e.g. LLL Coulomb) interactions. While for simple Jain series such numerical evidence is strong, generalizing the CF theory to more complicated FQH states can be more speculative. In some cases the LLL projection will cause the entire wavefunction to vanish, which may not be easy to predict until the actual numerical calculation is done\cite{PhysRevB.88.205312}. On the other hand it is important to note that while it is in principle possible that the LLL projection may change the topological properties of the FQH states, that is believed not to be the case for the Jain series and many other more exotic topological phases\cite{arxiv.2205.11214}.

It is useful to first think about if the \emph{unprojected} CF wavefunctions are relevant to the physics of the FQH states they are supposed to describe. There have been attempts in finding model Hamiltonians for the unprojected CF wavefunctions. Such approaches either require the projection of the Hilbert space into several low-lying LLs\cite{PhysRevB.44.8395,PhysRevLett.124.196803}, or into the subspace where the number of particles in each LL is conserved\cite{PhysRevLett.126.136601}. Both projections have to be implemented by hand, and cannot be realised by taking the limit of the cyclotron energy going to zero (they are thus toy models not easily associated to experiments). It is also not feasible to check if these model Hamiltonians are adiabatically connected to the realistic ones with large cyclotron energy, which in practice is the necessary ingredient for realising any FQH phases. Note that in principle such checking requires finite size scaling, as for any finite systems there is level repulsion; for example the Pfaffian and anti-Pfaffian phases are adiabatically connected for any finite systems on the torus, even though we know they are topologically distinct. Unfortunately finite size scaling for the adiabatic gap is generally impossible due to the small system sizes that are numerically accessible. 

From a phenomenological point of view, the CF theory does not have to insist on the LLL projection, as long as the unprojected CF wavefunctions can describe the essential physics of the FQHE (e.g. the filling factor, the topological shift, the edge physics, etc.) and explain the experimental data. Indeed in many cases the effective field theories are constructed from the unprojected CF theory only\cite{blok1990effective,PhysRevB.46.2290}. The danger of this approach is two-fold, given that the physics of FQHE entirely arises from the Hilbert space of a single Landau level, and all universal properties should agree with the limiting case when the magnetic field goes to infinity (or the effective mass goes to zero). First of all, it is not guaranteed that the projected CF wavefunctions have the same physics as the unprojected ones; the predictive power of the CF theory thus almost entirely rely on finite size numerical checking. One should note finite size numerical analysis by no means \emph{always} predict the same behaviours in the thermodynamic limit. For FQH topological phases, just looking at the ground state wavefunction overlap for finite systems is not sufficient; one has to also analyse the incompressibility gap, the overlaps of the quasiholes, as well as the counting and the bandwidth of the quasiholes. All these properties depend very sensitively on the details of the electron-electron interaction especially when the accessible system sizes are small, and they need to be taken into account when making predictions with the effective field theories.

Secondly, it is desirable to have a more microscopic derivation of the CF theory, to understand if the unprojected CF wavefunctions (and thus the LLL projection itself) are fundamentally necessary, or if they are just auxiliary numerical tools convenient for some large scale numerical computations. The LLL projection is a non-unitary process that does not have any tuning parameters. This implies that the unprojected CF wavefunctions do not really capture the physics of the LL mixing, since the mixing is continuously controlled by the details of the Hamiltonian (i.e. the relative strength between the interaction and kinetic energy). For cases where the higher LL components of the unprojected wavefunctions are small (e.g. the Jain series, where the overlap with the LLL projected wavefunction is high), then from a topological point of view the projected and unprojected CF wavefunctions are equivalent or adiabatically connected. In other cases, even if the unprojected wavefunctions are still conjectured to correspond to the projected ones, their physical properties (e.g. the variational energy) can be very different due to the large component in higher LLs.

One can thus highlight some of the undesirable properties of the LLL projection. It produces many-body wavefunctions that are not linearly independent and sometimes vanishing, and the resulting states \emph{do not} have physical or model Hamiltonians\cite{PhysRevB.98.235139} such that they are the \emph{exact} eigenstates. It is highly specific to the LLL, while in contrast the same topological phases can be realised in any LL with the proper microscopic Hamiltonians. The LLL projection is also not exactly compatible with the particle-hole conjugation, a well-defined unitary operation within a single LL. For example, the CF wavefunction of the Laughlin phase at $\nu=1/3$ is not the exact PH conjugate of the CF wavefunction of the anti-Laughlin phase at $\nu=2/3$, though it should be noted the pair of CF wavefunctions are approximately PH conjugate to a very high level of accuracy from numerical computations. Probably most importantly the projection is a non-unitary operation, that prevents us from treating the composite fermions as rigorous microscopic objects, beyond a phenomenological description. While there is very little doubt that these concerns may not be important for the simplest Abelian Jain series in the LLL, a careful understanding of its physical implications (and its necessity) may be important for applying CF theories to other more interesting (and more exotic) FQH phases. 

\textit{Relationship to other theories--}
Perhaps the more interesting issues concern with the various different methods in understanding the FQH effect. The full microscopic Hamiltonian for electrons in a quantum Hall system is given by:
\begin{eqnarray}\label{full}
\hat H_{\text{full}}=\hat H_{\text{kinetic}}+\hat H_{\text{int}}
\end{eqnarray}
where the first term on the RHS, the single particle kinetic energy Hamiltonian that gives the Landau levels, is the dominant energy scale. Thus at low temperature, the relevant dynamics is within a single LL, which is a strongly interacting system with a constant kinetic energy. With no usual perturbative techniques applicable, many new mathematical tools are developed in trying to characterise such systems. In addition to the CF theory, the FQH states were first studied with model wavefunctions and the model Hamiltonians\cite{laughlin1983anomalous,prangegirvin}. Later on the two-body model Hamiltonians were generalised to few-body model Hamiltonians in the form of generalised pseudopotentials\cite{PhysRevB.75.075318}. These are interaction Hamiltonians that effectively project into certain relative angular momentum sectors of a cluster of electrons, and they offer the \emph{minimal} models for both abelian and non-abelian topological phases. It was also discovered that the model wavefunctions of some FQH phases can be reinterpreted as conformal blocks of certain conformal field theories\cite{cristofano1993topological,moore1991nonabelions,RevModPhys.89.025005}, or as Jack polynomials\cite{PhysRevLett.100.246802} and their generalisations\cite{PhysRevLett.108.256807,PhysRevLett.112.026804,PhysRevB.100.241302}. The null spaces of the model Hamiltonians have conformal symmetry leading to the bulk-edge correspondence and universal edge dynamics of the quantum Hall fluids\cite{PhysRevB.79.245304,yang2021gaffnian,PhysRevLett.101.246806}. Effective theory approaches including the conformal field theory (CFT) and the topological field theories (TFT) become powerful tools in conjecturing about the nature and the properties of many FQH states. It is, however, difficult to justify the rigour and applicability of these effective field theory descriptions from the microscopic picture.

In principle, while the same physics can be understood from different perspectives, these different perspectives can be reconciled among one another in a consistent and unambiguous manner. This has yet to be the case for the FQH effect. It is intriguing why we cannot find local exact quantum Hamiltonians for all of the CF wavefunctions except for the Laughlin states. The failure in this implies some aspects of these CF wavefunctions are not physically relevant (since they may not be physically realisable even in principle), and they may not be the simplest ways of describing these FQH phases (e.g. the Jain series). An illustrative example is the anti-Laughlin states, or the particle-hole conjugates of the Laughlin states that have simple well-defined model wavefunctions and model Hamiltonians. The CF wavefunctions from reverse flux attachment and the LLL projection, however, are more complicated and with no known local Hamiltonians. They have very high overlap with the known model wavefunctions, and thus captures all the essential physics of the anti-Laughlin topological phases. Nevertheless, the differences between these CF wavefunctions and the exact anti-Laughlin wavefunctions are in principle physically irrelevant, and it is desirable for such differences to be eliminated within the CF theory.

The popular effective topological field theory (TFT) description is usually based on the IQH description of CFs or partons before the LLL projection. In principle, such TFT describes a rather different topological system not within a single LL. It is not easy to justify numerically for some CF states that the topological nature remains the same after the LLL projection. This is especially true for the quasihole manifold that determines the central charge of the edge theory and the braiding (abelian v.s. non-abelian) of the quasiholes. The predictions from the effective theories would require the quasihole bandwidth to be much smaller than the temperature, which may not be the case in the presence of realistic interactions\cite{yang2019effective,PhysRevB.80.205301}. Numerical analysis also shows even for the simplest FQH states, the LLL projection introduces missing states\cite{PhysRevB.88.205312}, non-trivial dynamics (e.g. non-zero dispersion and bandwidth of the quasihole excitations\cite{PhysRevB.88.205312}), CF level mixing\cite{PhysRevB.96.245142}, etc. All these are experimentally relevant for the robustness of the topological properties predicted by the effective TFT. 
 
 The relationship between the model wavefunctions and the conformal blocks lead to the conjecture that dynamical properties of certain FQH states can be governed by conformal symmetry in two dimensions. For example the Gaffnian and the Haffnian phases are believed to be gapless, given the associated CFT are non-unitary and/or irrational\cite{PhysRevB.79.045308,PhysRevB.83.241302}. Interestingly, some CFT related FQH states have very high overlap with the Jain series, and the corresponding ground states have identical topological indices\cite{simon2007construction,PhysRevB.81.121301,PhysRevB.100.241302}. From a dynamical point of view, however, the CFT related states are believed to be gapless and non-abelian, while the Jain series are believed to be gapped and abelian. It is nevertheless important to understand the fundamental relationship between the CFT formalism and the CF theory, given that most of the arguments are based on effective theories and numerical evidence\cite{kang2017neutral,jolicoeur2014absence}. 

\section{The objectives}
In this paper, we propose a general fermionization scheme that closely mirrors the spirit of the CF theory, yet is distinctive in several important microscopic aspects. The goal is to establish a more complete microscopic understanding of the underlying postulates of the CF theory, as well as to systematically examine the connections between the CF theory, the effective CFT and TFT descriptions, and the model Hamiltonians. At a more practical level, we propose new numerical methods in computing the properties of the FQH liquids, and point to possible sources where the experimental measurements can differ from the predictions of the CF theory, especially for the more exotic FQH phases requiring stringent constraints on different energy scales. The latter is made possible with the construction of model Hamiltonians within a single LL for a large number of CF-based FQH states, with which we can compare with the realistic Hamiltonians in various experiments. 

Explicitly, we establish that the composite fermions as emergent particles can be rigorously constructed as a microscopic basis obtained from a unitary transformation of the electron basis. In this new framework, CF wavefunctions are constructed directly within a single LL, without the need of the LLL projection. We argue these CF wavefunctions are topologically equivalent to the conventional CF wavefunctions for the Jain series, and the difference between them are unimportant non-universal physics that is irrelevant given the unpredictable microscopic details in experiments. Also unlike the conventional cases, the CF wavefunctions proposed here have exact model Hamiltonians. The concepts of ``$\Lambda$" levels (referred to as CF levels in this work) and the interaction between CFs are no longer phenomenological from numerical analysis, but now can be analytically defined with explicit microscopic Hamiltonians. 

The model Hamiltonians proposed here with the familiar pseudopotential formalism are also analytical tools to implement the fundamental postulates of the CF theory \emph{exactly}, that we are mapping a strongly interacting FQH phase of electrons to a \emph{non-interacting} IQH phase of composite fermions. For such mapping to be exact, we need the degenerate lowest energy eigenstates (e.g. the ground state and the quasihole states) of the Hamiltonians to be product states of the CFs,  much like the case of the IQH from the single particle kinetic energy. The CFs need to be non-interacting and non-dispersive, and the Hamiltonians cannot mix different CF levels, analogous to the kinetic energy not mixing the Landau levels. Note that with the Coulomb interactions, all these properties are only approximately satisfied even for the simple Jain series. With the model Hamiltonians as references, we can thus identify which part of the realistic interactions are ``perturbations" to the model Hamiltonians, and such perturbations induce interaction, dispersion and CF level mixing in real systems. 

We apply the fermionization scheme to the well-known Jain series and the composite fermi liquid, as well as their particle-hole conjugate states. For all these cases, the model Hamiltonians within a single LL can be explicitly constructed in principle, of which the ground states and the quasihole states are exact degenerate (i.e. zero energy after the proper constant energy shift with respect to the chemical potential if needed) eigenstates. The microscopic model wavefunctions are completely within a single LL by construction, and we hope to argue using these examples that all FQH states that can be understood via the CFs should be constructed in the same manner. Indeed, the fermionization scheme allows us to take any FQH states of electrons, and replace the electrons with any type of CFs in a rigorous way via a unitary transformation (together with their model Hamiltonians). In this way, a large family of FQH states can be constructed, going beyond the Jain series to include states arising from strong interaction between CFs. While the ``same" FQH states with different types of CFs (of which the electrons are one special case) occur at different filling factors, they are inherently physically equivalent. We will show in details on how starting from the principle Read-Rezayi series, the filling factors of the family of the FQH states constructed from the principle series has a fractal distribution, allowing us to understand a large number of FQH states in a systematic and unified manner. 

The organization of the paper will be as follows: In Sec.\ref{scheme}, we show explicitly the fermionization scheme on how the CFs emerge as a unitary transformation from the electron basis that can be naturally defined within the null spaces of model electron-electron interactions, and we list a set of criteria that needs to be satisfied for the mapping between the FQH of electrons to the IQH of CFs we have constructed. In Sec.\ref{jain} we apply the fermionization scheme to the familiar Jain series, in particular showing that the FQH states of the series we have constructed do not require the LLL projection and they all have exact model Hamiltonians. One example is an exact model Hamiltonian for the Jain $\nu=2/5$ state, which is incompressible, Abelian, and the ground state is \emph{identical} to the Gaffnian model state. Moreover, the particle-hole (PH) conjugation of the CFs can be naturally defined, leading to the construction of the PH conjugate Jain series again satisfying the exact FQH to IQH mapping without the LLL projection. In Sec.\ref{cfl} we focus on the composite fermi liquid (CFL), proposing the exact model wavefunctions (and the model Hamiltonians) for this gapless phase. We discuss about the PH conjugation of the CFL, the microscopic calculation of its dynamical properties, as well as implications to the effective theories and experimental results. While we mostly use the CFL at $\nu=1/2$ as an example, the discussions in this section also apply to other CFL at $\nu=1/(2p), p>1$. In Sec.\ref{fractal}, we treat the electrons and the CFs on the equal footing and illustrate a fractal structure for the distribution of the filling factors on the real axis, corresponding to the Abelian and non-Abelian FQH states of CFs, all of them with model Hamiltonians that can be explicitly constructed. This gives a natural way of understanding many of the FQH states from strongly interacting composite fermions. In Sec.\ref{summary} We summarise the main results and discuss about the future outlooks.

\section{A general fermionization scheme}\label{scheme}

Just like we can have many different types of bosons in the bosonization scheme to describe the same FQH phase, we can also have different types of fermions when we fermionize the many-body states. The composite fermions are constructed phenomenologically by attaching each electron with an even number of magnetic fluxes\cite{jain1989composite,jain2007composite}. We will closely examine this process and start with a vacuum on the sphere with an \emph{initial} magnetic monopole strength $2\bm S$, which is fixed. Adding electrons to the vacuum does not change the number of orbitals, and each electron is a spinor in the LLL with total spin $\bm S$, independent of the number of electrons added. Now let us consider adding composite fermions (CFs), each with one electrons and $q$ fluxes (and $q$ is even). We denote such composite fermions as $\text{CF}_q$. Again we require each $\text{CF}_q$ to have a fixed spin $S$; for it to be a good particle quantum number, $S$ needs to be \emph{independent} of the number of $\text{CF}_q$ added, even though adding CFs changes both the number of electrons and the magnetic fluxes on the sphere. Unlike the case for electrons, here we do not require $S=\bm S$.

Let us first look at the case when one $\text{CF}_q$ is added to the vacuum, giving a state with one electron and $2\bm S+1+q$ orbitals. This electron thus is a spinor with total spin $S=\bm S+\frac{q}{2}$. Since this is a state with a single $\text{CF}_q$, we will treat this $\text{CF}_q$ as a fermion with $S=\bm S+\frac{q}{2}$ as well. A multi-$\text{CF}_q$ state with $k$ CFs should thus be represented as $|m_1,S;m_2,S;\cdots;m_k,S\rangle$. To achieve this in a well-defined way, we need to find the Hilbert space spanned by $|m_1,S;m_2,S;\cdots;m_k,S\rangle$, such that there is a unitary transformation between $|m_1,S;m_2,S;\cdots;m_k,S\rangle$ and the many-body wavefunctions in the electron basis in this Hilbert space. It is also straightforward to reveal the statistical properties of the $\text{CF}_q$ once the Hilbert space is defined\cite{PhysRevLett.127.126406}.

To define such a Hilbert space, note that a many-body state with $k$ CFs contains $k$ electrons and $2\bm S+1+kq$ orbitals. In the full Hilbert space the eigenstates of $\hat L^2$ within this subspace give the counting that does not indicate we have $k$ fermions, each with a spin $S$. Instead as expected, the counting corresponds to $k$ fermions (i.e. electrons) each with a spin of $\bm S+kq/2$ (note the $k$ dependence). Thus in the full Hilbert space, the $\text{CF}_q$ we defined are not really ``particles". It turns out (not surprisingly, since the conventional CF wavefunctions give exact Laughlin wavefunctions without LLL projection) that the only way for these spin $S$ particles to behave like fermions is to confine the Hilbert space to the null space of the following model Hamiltonian:
\begin{eqnarray}\label{hq}
\hat H_q=\sum_{i=1}^{q/2}\hat V^{\text{2bdy}}_{2i-1}
\end{eqnarray}
where $\hat V^{\text{2bdy}}_{i}$ is the $i^{\text{th}}$ two-body Haldane pseudopotential (PP)\cite{prangegirvin}, and $\hat H_q$ is thus the model Hamiltonian for the Laughlin phase at $\nu=1/\left(q+1\right)$. We denote the null space of $\hat H_q$ as $\mathcal H_q$. If we diagonalise $\hat L^2$ within $\mathcal H_q$ for $k$ electrons and $2\bm S+1+kq$ orbitals, the eigenstate counting corresponds exactly to $k$ fermions, each with spin $S$ that is \emph{independent} of $k$. We thus show that spin $S$ composite fermions with $q$ fluxes attached to one electron are emergent fermionic particles with microscopic interaction $\hat H_q$ between electrons. 

More specifically, let us look at states with $N_e$ electrons and $N_o$ orbitals in $\mathcal H_q$, which is the subspace that automatically imposes the constraint that $N_o\ge \left(q+1\right)N_e-q$. We also require $N_e\ge q$, and this will become apparent later on. All states in $\mathcal H_q$ can be organised as simultaneous eigenstates of $\hat L^2$ and $\hat L_z$, given as follows:
\begin{eqnarray}
&&|l_z,l,\alpha_{l_z,l}\rangle_e=\sum_\lambda c_{l_z,l_,\alpha_{l_z,l},\lambda}|m_\lambda\rangle_e\\
&&\hat L_z|l_z,l,\alpha_{l_z,l}\rangle_e=l_z|l_z,l,\alpha_{l_z,l}\rangle_e\\
&&\hat L^2|l_z,l,\alpha_{l_z,l}\rangle_e=l\left(l+1\right)|l_z,l,\alpha_{l_z,l}\rangle_e
\end{eqnarray}
where $\alpha_{l_z,l}$ labels the degeneracy of states with quantum numbers $l_z,l$, and $|m_\lambda\rangle_e$ are the monomials, or Slater determinants of electrons with $N_e$ electrons and $N_o$ orbitals. We now look at the full Hilbert space of $N_e$ fermions in 
$\tilde N_o=N_o-q\left(N_e-1\right)$ orbitals. Its Hilbert space dimension equals exactly with that of $\mathcal H_q$, so the states again can be organised as simultaneous eigenstates of $\hat L^2$ and $\hat L_z$:
\begin{eqnarray}
&&|l_z,l,\alpha_{l_z,l}\rangle_{\text{CF}}=\sum_\lambda d_{l_z,l_,\alpha_{l_z,l},\lambda}|m_\lambda\rangle_{\text{CF}}\\
&&\hat L_z|l_z,l,\alpha_{l_z,l}\rangle_{\text{CF}}=l_z|l_z,l,\alpha_{l_z,l}\rangle_{\text{CF}}\\
&&\hat L^2|l_z,l,\alpha_{l_z,l}\rangle_{\text{CF}}=l\left(l+1\right)|l_z,l,\alpha_{l_z,l}\rangle_{\text{CF}}
\end{eqnarray}
where $|m_\lambda\rangle_{\text{CF}}$ are the monomials, or Slater determinants of this new types of fermions with $N_e$ fermions and $\tilde N_o$ orbitals. The one-to-one mapping between $|l_z,l,\alpha_{l_z,l}\rangle_e$ and $|l_z,l,\alpha_{l_z,l}\rangle_{\text{CF}}$ thus allows us to define the following unitary transformation:
\begin{eqnarray}\label{unitary}
|m_\lambda\rangle_{\text{CF}}&&=\sum_{l_z,l,\alpha_{l_z,l}} \tilde d_{l_z,l_,\alpha_{l_z,l},\lambda}|l_z,l,\alpha_{l_z,l}\rangle_{\text{CF}}\nonumber\\
&&\coloneqq\sum_{l_z,l,\alpha_{l_z,l}} \tilde d_{l_z,l_,\alpha_{l_z,l},\lambda}|l_z,l,\alpha_{l_z,l}\rangle_e\nonumber\\
&&=\sum_{l_z,l,\alpha_{l_z,l},\lambda'} \tilde d_{l_z,l_,\alpha_{l_z,l},\lambda}c_{l_z,l_,\alpha_{l_z,l},\lambda'}|m_{\lambda'}\rangle_e
\end{eqnarray}
where each monomial in the CF basis is a linear combination of the Laughlin quasihole states, and the coefficients $\tilde d_{l_z,l_,\alpha_{l_z,l},\lambda},c_{l_z,l_,\alpha_{l_z,l},\lambda'}$ can be readily obtained from the $\hat L^2$ diagonalisation.

The null space of Eq.(\ref{hq}) is spanned by the familiar Laughlin states and their quasiholes. This is the only family of FQH states where the conventional CF construction is ``exact", in the sense the conventional CF wavefunctions agree with model wavefunctions exactly without the need of the LLL projection. Thus these are the only conventional CF wavefunctions where model Hamiltonians exist. What we show here is that a unitary transformation between the electrons and $\text{CF}_q$ is only possible for a specific truncated Hilbert space: the null space of Eq.(\ref{hq}). It is thus an emergent property, and a fermionization process that is only exact with a specific toy model electron-electron interaction, which is different from the realistic LLL Coulomb interaction. We will extend the same interpretation to other CF states in the following sections.

\textit{The FQH and IQH correspondence--}
The fundamental postulate of the CF theory is that the fractional quantum Hall states of the electrons can be mapped to the integer quantum Hall (IQH) states of a new type of fermions, the composite fermions. 
In many cases, the effective topological field theories (TFT) are constructed in the context of the \emph{IQH systems}, and they are used to describe the topological phases of the \emph{FQH systems} (e.g. single component fermions in a partially filled LL), with the assumption of the validity of the FQH to IQH mapping. It is thus important to better understand the FQH and IQH correspondence in more details. The prediction of the topological indices of the FQH phases from the effective field theories based on the CF construction requires such correspondence to be robust in the presence of the various different energy scales in realistic systems. 

It is necessary to emphasize again that in the effective topological field theory description, all energies are sent \emph{by hand} to either infinity or zero. Let $E_i$ be the energies are sent to infinity (e.g. the incompressibility gap that is responsible for the robustness of the Hall conductivity and the Hall viscosity), and $\epsilon_i$ be the energies that are sent to zero (e.g. the quasihole energy bandwidth that is responsible for the robustness of the central charge and the non-abelian braiding). Each of these energies emerge from the interaction between electrons in the realistic system, and generically they could all be finite. The necessary condition for the topological properties as predicted by the effective field theories to be robust is thus for $E_i\gg E_T\gg\epsilon_i$, where $E_T$ is the energy scale of temperature and disorder.

The IQH systems with an infinite magnetic field clearly satisfy the ideal conditions of the effective TFT, where the incompressibility gap is infinity and the quasihole bandwidth is strictly zero when the interaction effect is ignored. In the conventional CF theory, however, for practical purposes the Coulomb interaction between electrons is often used. With this microscopic interaction Hamiltonian (or similar Hamiltonians after taking into account of the sample thickness and LL mixing, etc), the FQH and IQH correspondence is not exact, at least at the quantitative level. Unlike electrons in IQH, the CFs are not non-interacting, though the effective interaction between CFs could be weaker than that between electrons. Thus states in a single CF level (in analogy to the LL) are not degenerate, and with mixing to other CF levels. While in many cases the ``imperfect" correspondence can be ignored because $E_i\gg E_T\gg\epsilon_i$ holds, it is not easy to know how general such correspondence is within the CF theory, or how to quantify the ``noises" coming from the LLL projection, especially for the more exotic topological phases.

We thus endeavour to formulate an exact correspondence between FQH and IQH, without requiring the process of the LLL projection. We define the nature of the ``exact correspondence" with the following attributes:
\begin{enumerate}[label={\bf A}\arabic*]
\item The FQH ground state and the quasihole states can be reinterpreted as \emph{product states} of CFs;
\item The ground state can be reinterpreted as a Slater determinant CFs in a \emph{fully filled} CF level;
\item The FQH ground states and quasihole states have \emph{exact} model Hamiltonians.
\end{enumerate}
The requirement of exact model Hamiltonians can also be made precise. Such Hamiltonians satisfy the following conditions:
\begin{enumerate}[label={\bf B}\arabic*]
\item The Hilbert space of the Hamiltonian is a single LL;
\item The FQH ground state and the quasihole states are exact eigenstates;
\item All states within a single CF level with the same number of CFs have the same energy, so there is no dispersion of or interaction between CFs;
\item There is no mixing between CFs within the CF level  and CFs outside of the CF level (in analogy to no LL mixing for electrons in IQH).
\end{enumerate}
All four conditions are needed for the correspondence between FQH and IQH to be exact. One should note that for IQH, the presence of Coulomb interaction will also lead to LL mixing. For Galilean invariant systems where the cyclotron energy scales with the magnetic field $B$ while the interaction scales with $\sqrt B$, the coupling between different LLs can be suppressed by taking $B\rightarrow\infty$. Thus for the exact model Hamiltonian on the FQH side, we should also allow the coupling between CFs within the CF level  and CFs outside of the CF level to be suppressed by tuning some parameters in the Hamiltonian to infinity. Accordingly, our generalised fermionization scheme should allow the microscopic construction of CFs satisfying all conditions listed above.

\section{The Jain series and the related CF states}\label{jain}
We have already shown for the Laughlin states, the $\text{CF}_q$ emerges as well-defined, non-interacting quasiparticles only with pseudopotential model Hamiltonians. We will first examine the Laughlin states in more details, followed by generalising the same construction to the Jain series, as well as particle-hole conjugate states both in terms of electrons and CFs.

\subsection{The Laughlin states}

Let us first show the exact correspondence between the FQH and IQH can be established for the Laughlin states. These states are the simplest in the sense that in the CF theory, the ground states and the quasihole states do not require the LLL projection. Having established that in the sub-Hilbert space of $\mathcal H_q$, the $\text{CF}_q$ are indeed fermions with spin $S=\bm S+\frac{q}{2}$, we have shown that $\text{CF}_q$ as the emergent topological objects result from short range interaction between electrons, corresponding to the well-known model Hamiltonians of the Laughlin states. 

This also naturally allows us to have a ``Slater determinant" state of $\text{CF}_q$, when all possible single $\text{CF}_q$ state (a total of $2S+1$ of them) are occupied. One can denote such a state as $|-S,S;-S+1,S;\cdots;S-1,S;S,S\rangle$ which in itself is a many-body, strongly entangled state in the \emph{electron basis}. It is easy to check for such fully filled ``Slater determinant" states, the relationship between electron number $N_e$ and the orbital number $N_o$ is given by $N_o=\left(q+1\right)N_e-q$, so $|-S,S;-S+1,S;\cdots;S-1,S;S,S\rangle$ is precisely the Laughlin ground state at filling factor $\nu=1/\left(q+1\right)$. We can interpret these ground states as product states in the CF basis with no CF orbital entanglement, a statement that can now be made precise mathematically.

When fewer than $2S+1$ $\text{CF}_q$ are added to the vacuum, we obtain the Laughlin quasihole states. The quasihole subspace, or indeed $\mathcal H_q$, are spanned by orthogonal product states of the $\text{CF}_q$. Thus $\text{CF}_q$ in $\mathcal H_q$ is completely analogous to the non-interacting electrons in the LLL: the Hilbert space is spanned by the Slater determinants of the corresponding particles. In particular, there is \emph{no interaction} between CFs, even though the electrons are strongly interacting with $\hat H_q$.

Thus all attributes listed previously are satisfied, if and only if the interaction between electrons is given by the model Hamiltonians of the Laughlin phases. The more realistic Coulomb interactions, on the other hand, will lead to interaction between $\text{CF}_q$ within the same CF level, as well as the mixing of $\text{CF}_q$ within and outside of the CF level. The quasihole states will thus no longer be degenerate. More precisely, the $\text{CF}_q$ are weakly interacting with the Coulomb interaction only in the sense that the leading pseudopotential components of the LLL Coulomb interaction are dominant, giving the incompressibility gap as the dominant energy scale as compared to the interaction between CFs. Taking the Laughlin state at $\nu=1/3$ as an example, we can write the effective interaction as:
\begin{eqnarray}
\hat V_{\text{eff}}=\hat V_1^{\text{2bdy}}+\delta \hat V
\end{eqnarray}
With the LLL Coulomb interaction we have $\hat V_{\text{eff}}=e^{-\frac{1}{2}\bm q^2}/\bm q$ in the momentum space (where $\vec q$ is the 2D momentum and $\bm q=|\vec q|$). Only $\delta \hat V$, which consists of two-body pseudopotentials $\hat V_{i>1}^{\text{2bdy}}$, leads to interaction between the $\text{CF}_2$. Thus $\text{CF}_2$ are weakly interacting in the sense that $||\delta \hat V||<||\hat V_1^{\text{2bdy}}||$. Also, $\delta\hat V$ leads to mixing between different CF levels, and such level mixing can only be suppressed by taking $||\hat V_1^{\text{2bdy}}||\rightarrow\infty$.  

\subsection{Adding Composite Fermions to filled CF bands}

After adding $2S+1$ $\text{CF}_q$ with $S=\bm S+\frac{q}{2}$, we have a completely filled CF band, and the resulting ``Slater determinant" state corresponds to the Laughlin ground state at $\nu=1/\left(q+1\right)$. We can no longer add $\text{CF}_q$ to this filled band within $\mathcal H_q$, which is the null space of the pseudopotential Hamiltonian $\hat H_q$ given by Eq.(\ref{hq}). This implies that further addition of $\text{CF}_q$ will come with an energy cost given by $\hat H_q$. In analogy to adding electrons to a filled LL (i.e. electrons in higher LLs have larger spins on the sphere), the additional $\text{CF}_q$ will also have a larger spin. Starting with a filled CF band, we let the additional $\text{CF}_q$ carry a total spin of $S=\bm S+\frac{q}{2}+1$. We thus need to find the proper condition for these $\text{CF}_q$ to behave like fermions carrying the same total spin, independent of the number of $\text{CF}_q$ added.

We again illustrate this with the specific case of $q=2$, so the filled CF band with $S=\bm S+1$ corresponds to the Laughlin ground state at $\nu=1/3$. We can denote this as the lowest CF level. Additional $\text{CF}_2$ carry the spin of $S=\bm S+2$. If we add $k$ such $\text{CF}_2$ (where the maximum possible value of $k$ is $2\bm S+5$), we are looking at the sector with $N_e=2\bm S+3+k$ and $N_o=6\bm S+7+2k$. Diagonalization of $\hat H_1$ gives a branch of low-lying states giving the right counting of $k$ fermions in each $\hat L^2$ sector. All states, however, are in the complement of the null space $\mathcal H_1$.

Given that these additional $\text{CF}_2$ each has a total spin of $S=\bm S+2$, there are in total $2\bm S+5$ single CF states, which forms a band that we denote as the second CF level. With $\hat H_1$, the $\text{CF}_2$ in the second CF level will interact with each other, and they will also interact with $\text{CF}_2$ in higher CF levels (which we will define rigorously later on). We thus need to modify the electron-electron interaction, so as to make the FQH and IQH correspondence exact. It turns out that for the $S=\bm S+2$ $\text{CF}_2$ to have the right counting of the fermions in the second CF level, they have to stay in the null space of the Gaffnian model Hamiltonian constructed from the three-body pseudopotentials\cite{PhysRevB.75.075318}:
\begin{eqnarray}
\hat H_G=\hat V_3^{\text{3bdy}}+\hat V_5^{\text{3bdy}}
\end{eqnarray}
This implies with the following Hamiltonian:
\begin{eqnarray}\label{hprime}
\hat H'=\lambda_1\hat H_1+\lambda_3\hat H_G
\end{eqnarray}
we can define the second CF level as part of the null space of $\hat H_G$, when the first CF level is completely filled. In this way, the $\text{CF}_2$ in the second CF level are again well-defined fermions, but $\hat H_1$ will induce interactions between $\text{CF}_2$ within the second CF level, as well as mixing with states outside of $\hat H_G$ null space, which we can naturally define as higher CF levels. If we take the limit of $\lambda_3/\lambda_1\rightarrow\infty$, which is analogous to the tuning of $B\rightarrow\infty$ for the IQH, the mixing between different CF levels will be suppressed.  With $\hat H'$, as long as the lowest CF level is completely filled, the finite energy states are spanned by product states, or Slater determinant states of $\text{CF}_2$ in the second CF level.
\begin{figure}
\begin{center}
\includegraphics[width=\linewidth]{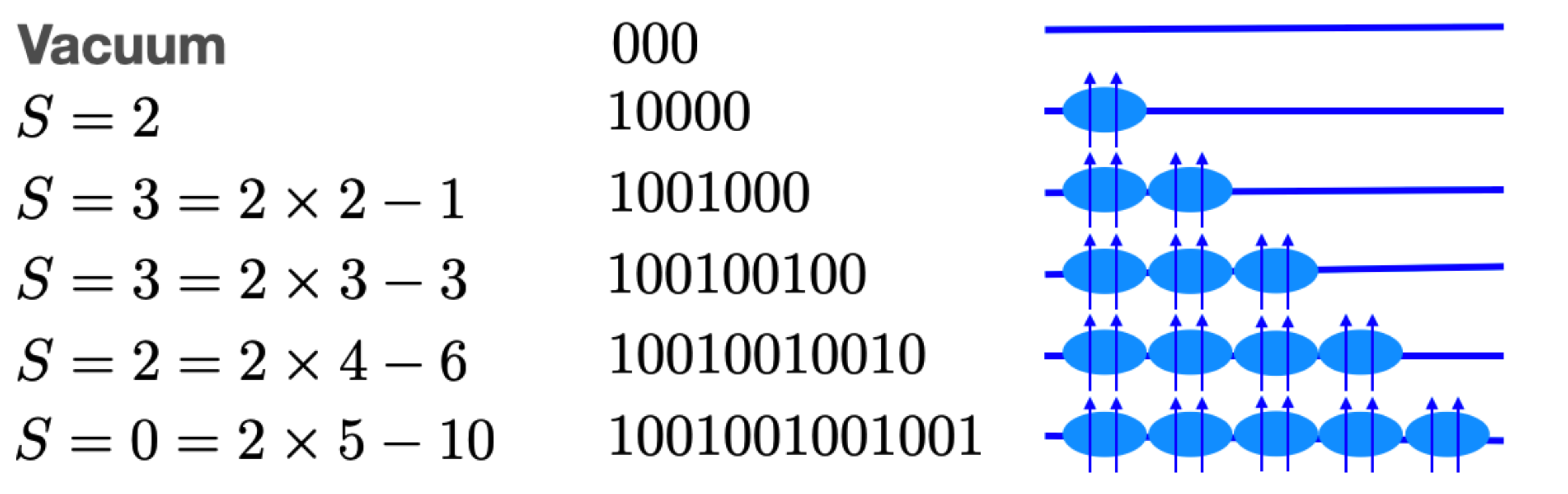}
\caption{Construction of the Laughlin state and Laughlin quasiholes, and the corresponding root configurations\cite{PhysRevLett.100.246802} (the middle column) of the highest weight states on the sphere, when the vacuum contains three magnetic fluxes. The left column gives total spin (or the total angular momentum on the sphere) of the multiple $\text{CF}_2$ states, obtained from the sum of individual $\text{CF}_2$ each with spin $\bm S+1=2$, subtracting the total \emph{relative} angular momentum. The right column gives the schematic representation of adding (at most) 5 $\text{CF}_2$ to a single CF level with three magnetic fluxes.}
\label{fig1}
\end{center}
\end{figure}

What we still need is for all these product states to be degenerate, if we would like to map the second CF level to the second LL in the IQH picture. To achieve that, let us look at the three-body pseudopotential $\hat V_6^{\text{3bdy}}$, noting that $\hat H_H=\hat H_G+\hat V_6^{\text{3bdy}}$ is the model Hamiltonian for the Haffnian state\cite{PhysRevB.83.241302}. The null space of $\hat H_H$ contains the null space of $\hat V_1$. Thus for the $\text{CF}_2$ in the lowest CF level, $\hat V_6^{\text{3bdy}}$ does not introduce any interaction between them. It will, however, introduce interactions between $\text{CF}_2$ in the second CF level, and also induce mixing between those $\text{CF}_2$ in higher CF levels. The latter, however, can be suppressed by taking $\lambda_3\rightarrow\infty$. We can thus construct the following Hamiltonian:
\begin{eqnarray}\label{hp}
\hat H''=\lambda_1\left(\hat V_1^{\text{2bdy}}+c_6\hat V_6^{\text{3bdy}}\right)+\lambda_3\hat H_G
\end{eqnarray}
Here $c_6$ is the parameter we can tune to minimize the bandwidth of the $\text{CF}_2$ in the second CF level. 

\begin{figure}
\begin{center}
\includegraphics[width=\linewidth]{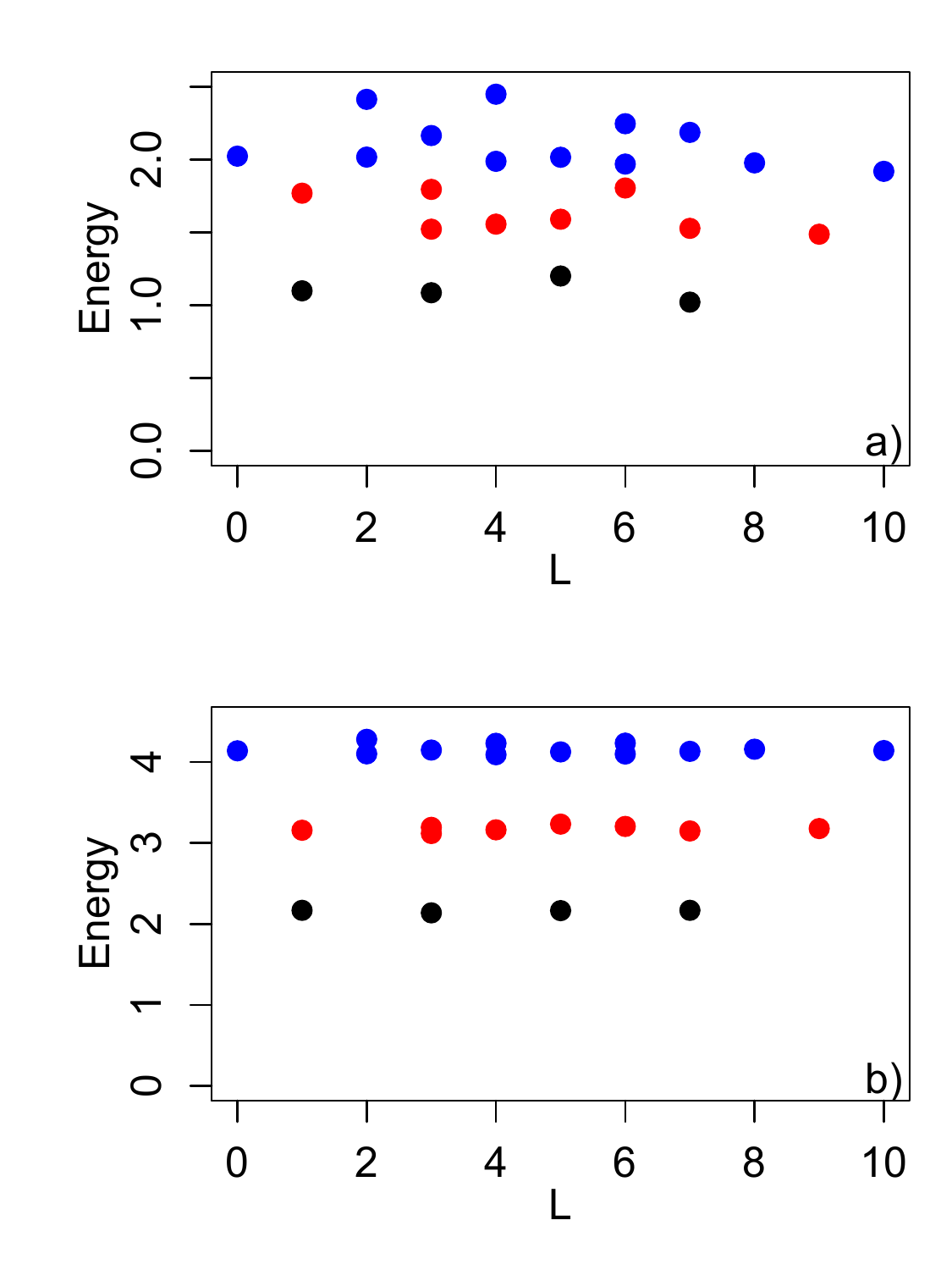}
\caption{The top panel is Eq.(\ref{hp}) with $c_6=0$, the bottom panel is Eq.(\ref{hp}) with $c_6=0.525$. The lowest CF level is completely filled with $9$ $\text{CF}_2$, and the second CF level contains 2 $\text{CF}_2$ (black plots), 3 $\text{CF}_2$ (red plot) and 4 $\text{CF}_2$ (blue plot). One can see the addition of $\hat V_6^{\text{3bdy}}$ can flatten the CF band and give better correspondence between the FQH and IQH.}
\label{fig2}
\end{center}
\end{figure}

From Fig.(\ref{fig2}) we can see the addition of $\hat V_6^{\text{3bdy}}$ can significantly flatten the band of CFs in the second CF level, at the same time giving a clear gap for neutral excitations with holes (or the absence of $\text{CF}_2$) in the lowest CF level. Further band flattening can be tuned by adding pseudopotential interactions $\hat V_k^{\text{n-bdy}}$, as long as its null space $\mathcal H_k^{\text{n-bdy}}$ satisfies the following condition:
\begin{eqnarray}\label{condition}
\mathcal H_L\subset \mathcal H_k^{\text{n-bdy}}\subset\mathcal H_G
\end{eqnarray}
where $\mathcal H_L$ is the null space of $\hat V_1^{\text{2bdy}}$, and $\mathcal H_G$ is the null space of the Gaffnian model Hamiltonian $\hat V_3^{\text{3bdy}}+\hat V_5^{\text{3bdy}}$. The three-body pseudopotentials satisfying Eq.(\ref{condition}) also include $\hat V_7^{\text{3bdy}}$ and $\hat V_8^{\text{3bdy}}$, and a general discussion about the hierarchy of the pseudopotential null spaces can be found in \cite{PhysRevB.105.035144,yang2020microscopic,multiplegraviton}. Since there are infinite number of suitable pseudopotentials if we include those involving more than three electrons, we conjecture that we can always find the right combination of them, which we denote as $\delta\hat V_1^{\text{n-bdy}}$, that allows us to make the band of $\text{CF}_2$ in the second CF level completely flat in the thermodynamic limit. The resulting Hamiltonian is given as follows:
\begin{eqnarray}\label{hh}
\hat H=\lambda_1\left(\hat V_1^{\text{2bdy}}+\delta\hat V_1^{\text{n-bdy}}\right)+\lambda_3\hat H_G
\end{eqnarray}
where the null space of $\delta\hat V_1^{\text{n-bdy}}$ contains $\mathcal H_L$ but is a proper subspace of $\mathcal H_G$. We propose this Hamiltonian should be the model Hamiltonian for the Jain state at the filling factor $\nu=2/5$. 

There are several important features for this construction. Here, the FQH and IQH correspondence is exact, in contrast to the original CF theory. The many-body wavefunctions of the ground state and quasihole states of the $\nu=2/5$ phase can all be understood as product states of $\text{CF}_2$  in the second CF level, and these states are constructed \emph{without} resorting to the projection into the LLL. The ground state of the $\nu=2/5$ phase is the Gaffnian model wavefunction, but the quasihole excitations are \emph{strictly abelian}, with exact correspondence to the IQH states. It is important to note that just like in the CF theory\cite{PhysRevB.80.205301,yang2019effective}, the quasiholes can only be created by removing CFs in the second CF level. The lowest CF level should be kept completely filled. The additional Gaffnian quasihole states responsible for the non-Abelian properties correspond to the removal of $\text{CF}_2$ in the lowest CF level, which cost a finite amount of energy due to the presence of $\hat V_1^{\text{2bdy}}$ in Eq.(\ref{hh}), and are gapped out at low temperature\cite{PhysRevB.80.205301,yang2019effective}.
\begin{figure}
\begin{center}
\includegraphics[width=\linewidth]{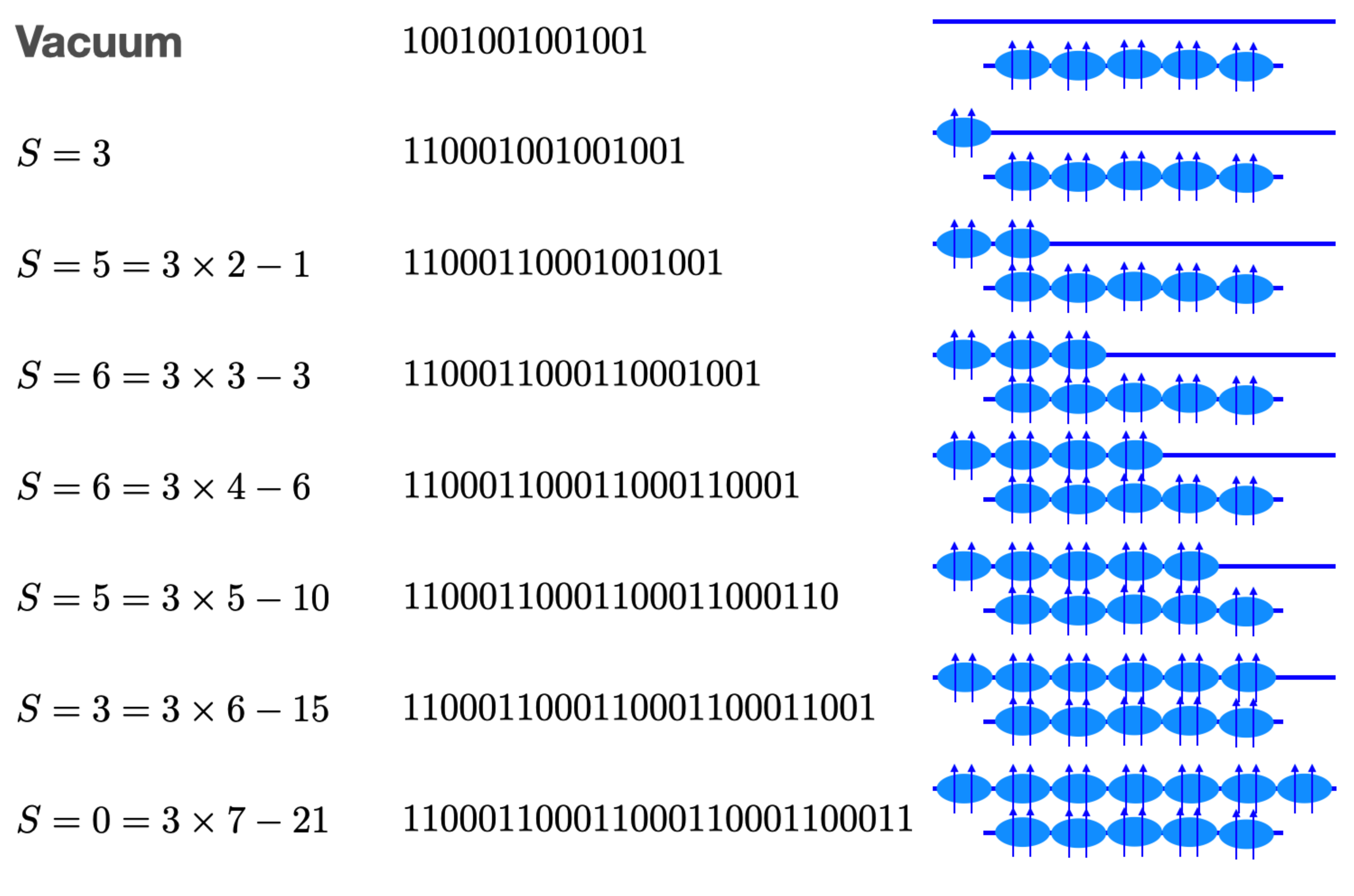}
\caption{Similar to Fig.(\ref{fig1}), here we show the construction of the Jain $\nu=2/5$ state and its quasiholes, with the corresponding root configuration. All these states are zero energy states of the Gaffnian model Hamiltonians.}
\label{fig3}
\end{center}
\end{figure}

There are arguments that the Gaffnian model Hamiltonian is gapless in the thermodynamic limit\cite{PhysRevB.79.045308,kang2017neutral,simon2007construction,jolicoeur2014absence}. With the model Hamiltonian we have constructed, however, it is clear the spectrum is gapped with finite $\lambda_1$, even though the ground state is the exact Gaffnian model state in the limit of $\lambda_3/\lambda_1\rightarrow\infty$. We thus show explicitly here a topological phase with Eq.(\ref{hh}), with all topological features identical to those claimed by the Jain $\nu=2/5$ phase and thus distinct from the Gaffnian phase, yet the ground state is identical to the Gaffnian model wavefunction\cite{yang2019effective}. One should note that all quasihole states of Eq.(\ref{hh}) are identical to the Gaffnian quasihole wavefunctions as well, if we only allow removal of the $\text{CF}_2$ in the second CF level. They form \emph{an Abelian subspace} of the entire Gaffnian quasihole manifold.

The scheme we propose can be extended to higher CF levels. When the lowest $N$ CF levels are fully occupied, additional $\text{CF}_2$ are fermions with total spin $S=\bm S+N+1$, which lives within the null space of a properly constructed $(N+2)-$body interaction. We conjecture it is always possible for model Hamiltonians to be constructed within the LLL, such that the FQH to IQH correspondence is exact for the entire Jain series at $\nu=n/\left(2n+1\right)$. The CF wavefunctions constructed this way are not identical to the conventional CF wavefunctions from the LLL projection. They however have very high wavefunction overlap for all numerically accessible system sizes. For example, for the next Jain state at $\nu=3/7$, we can construct a model Hamiltonian with 4-body pseudopotentials, leading to a model wavefunction related to the $S3$ conformal field theory\cite{PhysRevB.81.121301} or constructed from the local exclusion conditions\cite{PhysRevB.100.241302,yang2021elementary}. Such model wavefunctions have very high overlap with the conventional Jain state from the LLL projection, and the related quasiholes are contained in the null space of such model Hamiltonian. Thus similar to the Jain state at $\nu=2/5$, we can construct a model Hamiltonian for the Jain state at $\nu=3/7$ within the same LL with the spectrum that can be mapped to the IQH involving three CF levels.

All the arguments above can be generalised to the case of $q>2$, corresponding to the Jain series at the filling factors $\nu=n/\left(qn+1\right)$. While it is tedious to check for each state that the new approach here produces model wavefunctions that always have very high overlap with the wavefunctions from the conventional CF theory involving the LLL projection, they are indeed the case for the states we have checked. Though only limited numerics are performed to check that we can always construct model Hamiltonians giving non-interacting $\text{CF}_q$ with (almost) flat CF levels (especially for higher CF levels), its generality is likely true given that there are infinite number of parameters to tune, for there are infinite number of pseudopotentials with null spaces satisfying the condition analogous to Eq.(\ref{condition}) for all the Jain states. Independent of the conventional CF theory, the arguments here give a well-defined scheme for the construction of model wavefunctions and Hamiltonians for all Jain states within a single LL, from rigorously defined composite fermion product states.

\subsection{The particle-hole conjugation}

In a single LL, each FQH state at filling factor $\nu$ has a particle-hole (PH) conjugate partner at filling factor $1-\nu$, which is a distinct topological phase also with a well-defined microscopic model Hamiltonian\cite{levin2007particle,lee2007particle}. This PH conjugation is defined by the unitary transformation between electrons and holes. In the composite fermion theory, these states are constructed using the ingenious procedure of ``reverse flux attachment". This is a physically intuitive process, but with two caveats. Firstly, all states with reverse flux attachment requires the LLL projection, even for the PH conjugate state of the simplest Laughlin phases. Secondly, states constructed from reverse flux attachment are not exact PH conjugates of their partners at the microscopic level (even after the LLL projection), though numerically they tend to be very good approximations\cite{PhysRevB.93.235152} and are thus believed to describe the same topological phase (at least for the Jain states). As a consequence, there is no known exact model Hamiltonian even for the simplest states (e.g. the CF ``anti-Laughlin" state at $\nu=2p/\left(2p+1\right)$) from reverse flux attachment, though the model Hamiltonian for the actual ``anti-Laughlin" state at $\nu=2p/\left(2p+1\right)$ clearly should be the same as that of the Laughlin state at $\nu=1/\left(2p+1\right)$.

At the microscopic level, applying the PH conjugation to an electron many-body state is a straightforward procedure. In the second quantized language, the many-body wavefunction is a linear combination of the occupation basis, each representing a monomial or a Slater determinant state. The PH conjugate of this occupation basis can be simply obtained by switching an unoccupied orbital (denoted with ``0") to an occupied orbital (denoted with ``1"), and vice versa. For example, the Laughlin ground state $|\psi\rangle_L$ at $\nu=1/3$ with three electrons and seven orbitals, and its PH conjugate (the anti-Laughlin state at $\nu=2/3$) $|\psi\rangle_{AL}$, are given below:
\begin{eqnarray}
&&|\psi\rangle_L\sim |1001001\rangle-3|1000110\rangle-3|0110001\rangle\nonumber\\
&&\qquad\qquad+6|0101010\rangle-15|0011100\rangle\\
&&|\psi\rangle_{AL}\sim |0110110\rangle-3|0111001\rangle-3|1001110\rangle\nonumber\\
&&\qquad\qquad+6|1010101\rangle-15|1100011\rangle
\end{eqnarray}
Both states are the exact eigenstates of $\hat V_1^{\text{2bdy}}$. 

Let us for now call all the Jain states with $\nu<1/2$ as the usual FQH states, and their particle-hole conjugate states with $\nu>1/2$ as the PH FQH states. The PH conjugate of the vacuum for the FQH states is naturally the fully filled LL, which serves as the new vacuum (or the PH-vacuum) for the PH-FQH states. Similarly, since the $\text{CF}_q$ for the FQH states contains one electron and $q$ fluxes, its PH conjugate (which we call the $\text{PH-CF}_q$) consists of $q-1$ electrons and $q$ fluxes. This is analogous to the reverse flux attachment, or the composite fermion of ``holes", but here a $\text{PH-CF}_q$ is a microscopically well-defined fermion \emph{in a single LL}. It is easy to see that when we add $\text{PH-CF}_q$ to the PH-vacuum, they behave like $S=\bm S+\frac{q}{2}$ fermions in the null space of the PH conjugate of Eq.(\ref{hq}). Thus after adding $2S+1$ $\text{PH-CF}_q$, we get the ground state of the anti-Laughlin states at filling factor $\nu=q/\left(q+1\right)$, which again is a Slater determinant state of the $\text{PH-CF}_q$.
\begin{figure}
\begin{center}
\includegraphics[width=\linewidth]{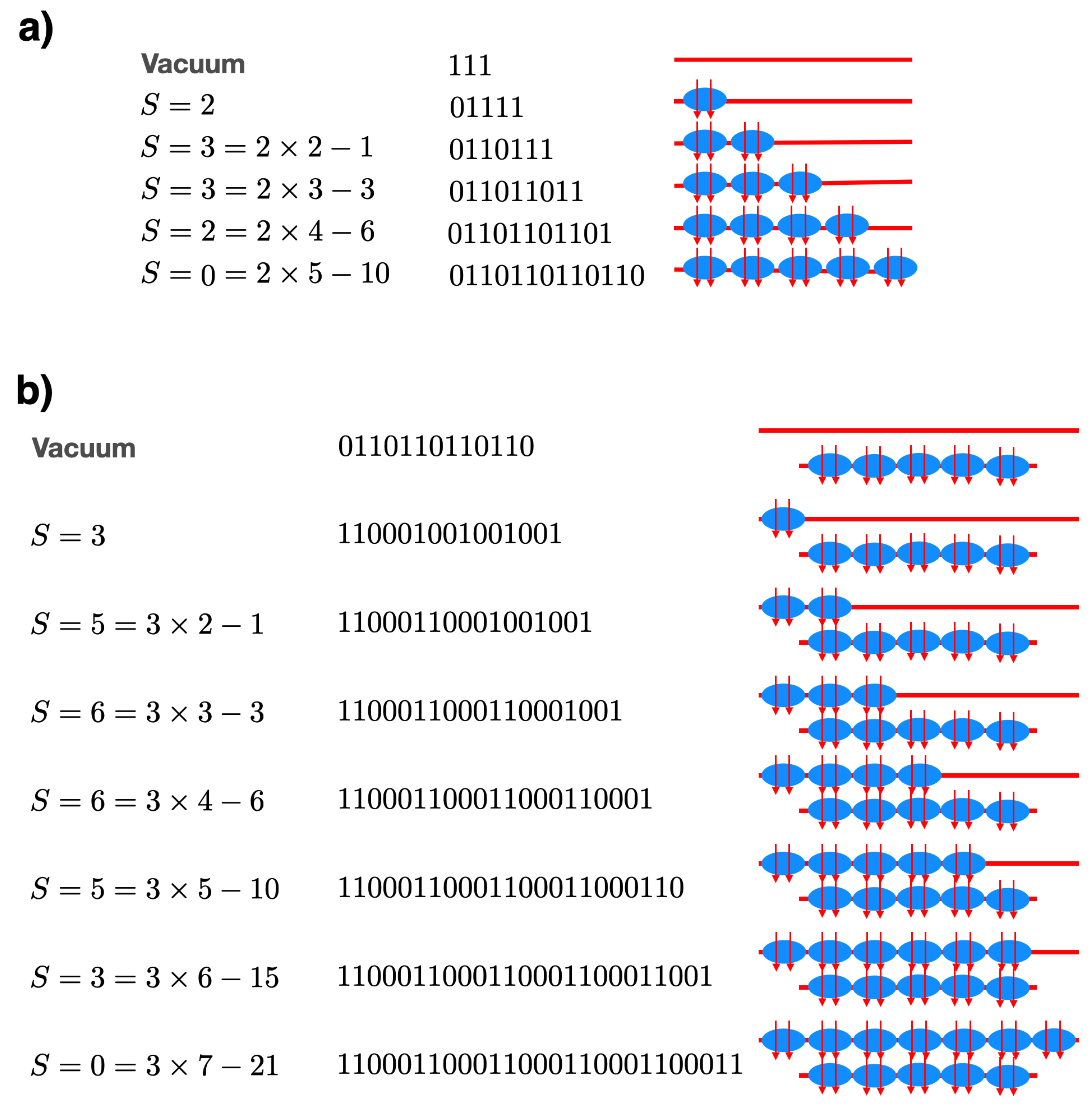}
\caption{a). The PH conjugate version of Fig.(\ref{fig1}), where $\text{PH-CF}_2$ are added to the vacuum of three magnetic fluxes to form the anti-Laughlin state at $\nu=2/3$. b). The PH conjugate version of Fig.(\ref{fig3}), where $\text{PH-CF}_2$ are added to the second CF level to form the PH conjugate of the Jain $\nu=2/5$ state.}
\label{fig4}
\end{center}
\end{figure}

Similarly, once the lowest PH-CF level is completely filled up with $2S+1$ $\text{PH-CF}_q$, further addition of $\text{PH-CF}_q$ will carry a spin of $S+1$, which can be defined using the PH conjugate of Eq.(\ref{hh}). Fully filling the second PH-CF level with $2S+3$ $\text{PH-CF}_q$ will give a product state of PH-CFs at $\nu=\left(2qn-1\right)/\left(2q+1\right)$ corresponding to the PH conjugate of the Jain state at $\nu=2/\left(2q+1\right)$. We propose all PH conjugate of the Jain series at $\nu=n/\left(qn+1\right)$ can be constructed in the same way. Note that in this way, not only are all the PH Jain states constructed without resorting to the LLL projection, they all have model Hamiltonians, which are simply the PH conjugate of the model Hamiltonians of the original Jain series.

\textit{The particle-hole conjugation of composite fermions--}
So far we have only discussed about the PH conjugation of electrons, leading to the transformation from $\text{CF}_q$ (one electron bound to $q$ fluxes) to $\text{PH-CF}_q$ ($q-1$ electrons bound to $q$ fluxes). Since PH conjugate is well-defined for any fermions, we can go beyond PH conjugation of electrons and construct FQH states with the PH conjugation that maps an orbital occupied with a $\text{CF}_q$ to an empty orbital, or vice versa. In this way, many more Jain states can be constructed without using the reverse flux attachment or the LLL projection. 

This can be illustrated with the Laughlin state at $\nu=1/5$. It is an IQH state of $\text{CF}_4$, each consisting of one electron and four magnetic fluxes, and the ground state corresponds to the fully filled CF level. On the other hand, we can also reinterpret it as the FQH state at filling factor $\nu=1/3$ of $\text{CF}_2$. In this case the CF level is partially filled, so we can take the particle-hole conjugate within this CF level. The resulting FQH state of the $\text{CF}_2$ at filling factor $\nu=2/3$ corresponds to the FQH state of electrons at $\nu=2/7$. This topological phase is well-known from the conventional CF theory, but here the state can be constructed without reverse attachment or LLL projection, but with an exact PH conjugation.

We would like to emphasise that every product state of $\text{CF}_q$ is a many-body wavefunction of electrons within a single LL. The PH conjugate of the CF product state within the CF level is thus another CF product state, which is also a well-defined many-body wavefunction of electrons within a single LL. In this way, we can write down the many-body wavefunction of the electron FQH state at $\nu=2/7$ explicitly. It is not microscopically identical to the state from the conventional CF theory, but it has an exact a model Hamiltonian. This microscopic Hamiltonian can be obtained by noting that for the $\nu=2/7$ phase, it is the anti-Laughlin phase at $\nu=2/3$ of $\text{CF}_2$. The model Hamiltonian for the electron state $\nu=2/3$ is well-known, and the model Hamiltonian for the $\text{CF}_2$ state at $\nu=2/3$ can thus be systematically derived from unitary transformation, which we will go in more details in Sec.~\ref{fractalmodelham}.

\section{The composite fermi liquid}\label{cfl}

One of the greatest achievements of the composite fermion theory is the model of the composite fermi liquid (CFL) for the compressible phase at filling factor $\nu=1/q$, where $q>0$ is even\cite{PhysRevB.47.7312,PhysRevLett.72.900}. Taken as the limit of the Jain series of $\nu=n/\left(qn\pm1\right)$ with $n\rightarrow\infty$, this is a gapless phase that is conjectured to be a fermi liquid of composite fermions with vanishing effective magnetic field. Using this model, the trial wavefunction can be constructed by flux attachment to the fermi liquid, followed by the LLL projection. A number of predictions of the composite fermi liquid theory are also supported by the experimental evidence\cite{PhysRevLett.114.236406,PhysRevB.100.041112,jain2007composite}.

Our first task is to see if the CFL wavefunction can be constructed without using the LLL projection, with the $\text{CF}_q$ we proposed in the previous section in a consistent manner. It is also worth noting that the traditional CFL wavefunction from flux attachment to a fermi liquid\cite{PhysRevLett.72.900} is not obvious from the limit of the Jain series at the microscopic level. Here we will take this limiting process seriously, which we will illustrate with the case of $q=2$. The CFL occurs at $\nu=1/2$, obtained from adding $\text{CF}_2$ to the vacuum with \emph{no effective magnetic fluxes} (in contrast to the example in Fig.(\ref{fig1}), where the vacuum has three effective magnetic fluxes). The lowest CF level thus has two single $\text{CF}_2$ states, with each $\text{CF}_2$ carrying the total spin of $1/2$. We can thus keep adding $\text{CF}_2$, filling up the consecutive CF levels. Note that no matter how many $\text{CF}_2$ are added, the filling factor is kept fixed at $\nu=1/2$. Moreover, when $N$ lowest CF levels are completely filled, we obtain the wavefunction which is \emph{a special ground state} of the Jain series at $\nu=N/\left(2N+1\right)$, with the number of electrons $N_e=N\left(N+1\right)$. This wavefunction has the filling factor at exactly $\nu=1/2$ at any value of $N$. One can also add or remove $\text{CF}_2$  from this special state, without changing the filling factor. 

The important message here is that if we start with the proper vaccum, the CFL ground state is well-defined for any finite systems, when a finite number of CF levels are completely filled, and the top CF level is partially filled. Thus topologically they are no different from the Jain states (ground states or quasihole states), with the additional constraint that the filling factor is exactly $\nu=1/2$. We thus also have a model Hamiltonian for the CFL as proposed in the previous section, leading to the microscopic wavefunctions that are automatically within a single LL. There is a root configuration of the CFL state with $N$ completely filled CF levels, given as follows:
\begin{eqnarray}\label{cfroot}
\underbrace{\overbrace{11\cdots 11}^N\overbrace{000\cdots 00}^{N+1}\cdots\overbrace{11\cdots 11}^N\overbrace{000\cdots 00}^{N+1}}_N\overbrace{11\cdots 11}^N
\end{eqnarray}
where all monomial basis of the CFL are squeezed\cite{PhysRevLett.100.246802} from the root configuration. For $N=1$ we have the Laughlin state at $\nu=1/3$ with two electrons, and for $N=2$ we have the Gaffnian state with six electrons, etc. All quasihole states created by removing CFs from Eq.(\ref{cfroot}) are also CFL states, with their corresponding root configurations satisfying no more than $N$ electrons in any $N+1$ consecutive orbitals. There is thus no fundamental difference between the CFL states and the Jain states describing the gapped FQH states, as they should be, though this was not explicit or apparent in the conventional construction. For small systems (and thus small $N$), the CFL states are gapped with short range interaction (e.g. $\hat V_{\text{LLL}}$). The gap however is expected to vanish in the limit $N\rightarrow\infty$, as we will discuss in Sec.~\ref{dynamics}. 
\begin{figure}
\begin{center}
\includegraphics[width=\linewidth]{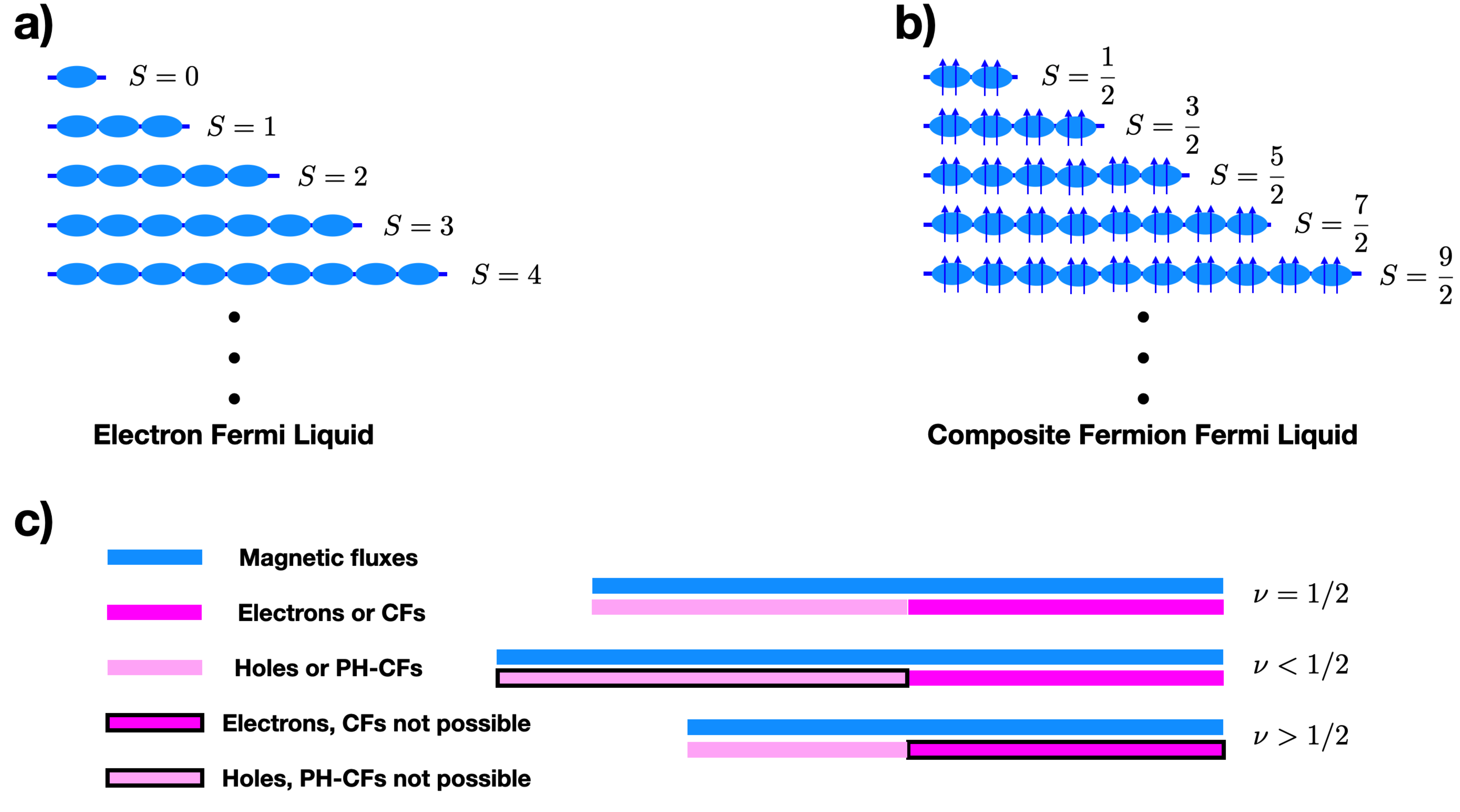}
\caption{a). For electrons, a fermi liquid on the sphere comes from adding electrons to the sphere without any magnetic field; b). similarly the CFL on the sphere at $\nu=1/2$ is obtained by adding $\text{CF}_2$  to the sphere with no effective magnetic field. c). Adding $\text{CF}_2$ to the vacuum can never bring the filling factor beyond $\nu=1/2$. When $\nu>1/2$, the physics of CFL can only be described by the $\text{PH-CF}_2$, which gives the particle density for the quantum fluid.}
\label{fig5}
\end{center}
\end{figure}

It is not possible to numerically construct such model states exactly for large value of $N$. We can, however, just diagonalise a short range interaction (e.g. $\hat V_1$ or $\hat V_{\text{LLL}}$) within the basis squeezed from the root configuration. The ground state in the truncated Hilbert space has extremely high overlap with the exact ground state from the full Hilbert space, indicating the CFL states we constructed indeed describe the physics at $\nu=1/2$ in the LLL, and they have high overlap with the CFL constructed conventionally with the LLL projection\cite{PhysRevLett.126.076604}. Our construction also shows the CFL states are highly PH symmetric, though in principle they do not have the \emph{exact} PH symmetry even in the thermodynamic limit.

\subsection{The emergence of the fermi surface}

The fermi surface of the $\text{CF}_2$ at $\nu=1/2$ is defined by the fermi wave vector $k_F$, which is an intrinsic property of the CFL wavefunction independent of the dispersion of the single CF. We show here that $k_F$ can be readily computed from the CFL wavefunction given by Eq.(\ref{cfroot}), which agrees with the numerical value computed from the conventional CFL wavefunction, or the value deduced from flux attachment to a fermi liquid\cite{PhysRevB.96.235102,PhysRevLett.115.186805}. The computation here is \emph{exact} in the thermodynamic limit.

From the fundamental definition of the fermi surface, $k_F$ is the smallest wave vector of a single $\text{CF}_2$ that can be added to a CFL. Starting with a rotationally invariant CFL with fully filled $N$ CF levels given by Eq.(\ref{cfroot}), an additional $\text{CF}_2$ to the next CF level carries a total spin $S=\frac{1}{2}+N$. Since the total number of magnetic fluxes of the CFL is $N_o=2\left(N^2+N+1\right)$, the radius of the sphere is given by $R=\sqrt{N^2+N+1}l_B$, where $l_B$ is the magnetic length. Thus the wave vector of the additional CF, which is the fermi wave vector, is given by:
\begin{eqnarray}
k_F=\lim_{N\rightarrow\infty}\frac{S}{R}=l_B^{-1}=\sqrt{4\pi n_{\text{CF}}}
\end{eqnarray}
where $n_{\text{CF}}=N_e/\left(4\pi R^2\right)$ is the CF density at $\nu=1/2$, with $N_e$ the number of electrons (or $\text{CF}_2$). For $\text{CF}_2$ constructed from flux attachment to electrons, the CF density is naturally the electron density. Similarly, for $\text{PH-CF}_2$ obtained from flux attachment to holes, its density is given by the \emph{hole density} that determines the fermi wave vector\cite{PhysRevLett.114.236406}. Here $k_F$ is determined by Eq.(\ref{cfroot}) describing \emph{non-interacting} $\text{CF}_2$ from interacting electrons, but since we have established a rigorous mapping between $\text{CF}_2$ and electrons, even with interacting $\text{CF}_2$, $k_F$ is universal due to the Luttinger theorem\cite{PhysRevB.96.235102,PhysRev.118.1417,PhysRev.119.1153}, as long as we are still in the fermi liquid phase. More importantly, $k_F$ is invariant to the specific form of the effective single CF dispersion, unlike the CF fermi energy. Such emergent dispersion depends on the details of the electron-electron interaction and is thus non-universal as we will discuss next. 

\subsection{The dynamics near the fermi surface}\label{dynamics}

Whether or not the CFL states are the proper gapless ground state at $\nu=1/2$ of course is entirely dependent on the electron-electron interaction. There is strong numerical evidence that the CFL states have very high overlap\cite{PhysRevLett.126.076604} with the exact ground states of short range (e.g $\hat V_1$ or $\hat V_{\text{LLL}}$) interactions. The same interactions also show the gap of the Jain series $\Delta_N$ to scale as $\Delta_N\sim1/N$\cite{PhysRevB.105.205147,Park_1999}, agreeing with the assumption that the CFs for the Jain state at $\nu=1/\left(2N+1\right)$ experience an effective magnetic field $B^*=B/\left(2N+1\right)$. This effective magnetic field is given by the number of magnetic fluxes of the vacuum. It is however important to note that $\Delta_N$ is numerically computed from finite systems where $N$ is small, and this is a point we will come back to later. It is also numerically shown that for short range interactions, at the fixed filling factor $\nu=1/\left(2N+1\right)$, the single $\text{CF}_2$ excitations are more or less equally spaced\cite{PhysRevB.93.235152,PhysRevB.64.081302} with energy spacing proportional to $B^*$. All these numerical evidence suggests that the CF levels are equally spaced with short range interactions. It is thus believed there is an emergent Galilean invariance for the $\text{CF}_2$ (or $\text{CF}_q$ in general), from which we can properly define an effective mass for the $\text{CF}_2$. Note that for electrons, only Galilean invariant systems (i.e. quadratic dispersion) leads to equally spaced Landau levels. 

It is important to point out, however, in principle all these are dynamical behaviours of the $\text{CF}_2$ in the presence of an effective magnetic field, which are necessary but \emph{not sufficient conditions} for the formation of CFL with the quadratic dispersion at $\nu=1/2$, when the effective magnetic field vanishes. They are also numerical evidence of simple interaction models for finite system sizes, and it is not clear how robust equal energy spacing for CF levels is in the thermodynamic limit, and against sample thickness, LL mixing and other factors in realistic experiments. For a more detailed analysis, we look at the CFL with $N$ fully filled CF levels. The exact expression of the fermi wave vector is given by:
\begin{eqnarray}\label{kf}
k_{F,N}=l_B^{-1}\frac{1+2N}{2\sqrt{N^2+N+1}}
\end{eqnarray}
which is the momentum of the $\text{CF}_2$ added to the $\left(N+1\right)^{\text{th}}$ CF level. Note that in contrast to the addition of one electron to the fermi surface at a fixed area that alters the electron density, here an addition of a $\text{CF}_2$ to the CFL fermi surface keeps the density fixed but increases the area (i.e. each $\text{CF}_2$ contains two magnetic fluxes). Let $E_N,\tilde E_{N+n,1}$ be the energy of the CFL before and after the addition of one $\text{CF}_2$ to the $\left(N+n\right)^{\text{th}}$ CF level, so $\Delta_{N,n}=\tilde E_{N+n,1}-E_N$ is a charge gap of the CFL, or the Jain state at $\nu=N/\left(2N+1\right)$. To get the fermi energy, we need to rescale the magnetic field to keep the area fixed, with the magnetic length $l_B\rightarrow l_B\sqrt{\left(N^2+N\right)/\left(N^2+N+1\right)}$. Since the energy of the CFL fundamentally comes from the Coulomb interaction between electrons which is inversely proportional to the magnetic length, after rescaling it gives us $E_{N+n,1}=\tilde E_{N+n,1}\sqrt{\left(N^2+N+1\right)/\left(N^2+N\right)}$. We thus obtain the following expression for the fermi energy, when the additional $\text{CF}_2$ is added to the $\left(N+1\right)^{\text{th}}$ CF level:
\begin{eqnarray}\label{ef}
E_F=E_{N+1,1}-E_N=\Delta_{N,1}+\frac{1}{2}\epsilon_N+O(N^{-1})
\end{eqnarray}
where $\epsilon_N=\tilde E_{N+1,1}/\left(N^2+N+1\right)$ is the energy density (per electron or $\text{CF}_2$).

All physically relevant measurements explore the dynamics near the fermi surface, and the scaling of $\Delta_N,\epsilon_N$ with respect to the system size completely determines the emergent single particle behaviours of the $\text{CF}_2$. The two quantities are also readily computed by numerics. The single particle dispersion of the $\text{CF}_2$ can be complicated and non-universal, just like band dispersions in crystals. Let us assume a generic dispersion to the leading order with $E_k=\alpha k^\beta$. We thus have the following relationships:
\begin{eqnarray}\label{dispersion}
E_F=\alpha k_F^\beta
\end{eqnarray}
Substitution of $E_F$ and $k_F$ into Eq.(\ref{ef}) gives us in the limit $\lim_{N\rightarrow\infty}\epsilon_N=\tilde \epsilon=2\alpha l_B^{-\beta}$, where $l_B$ comes from the magnetic field before rescaling, or before the addition of the $\text{CF}_2$. Remember that the CFL is gapless so $\lim_{N\rightarrow\infty}\Delta_{N,1}=0$. At the fermi surface, we also have the following fermi velocity:
\begin{eqnarray}\label{vf}
v_F=\frac{\partial E_k}{\partial k}\Bigr|_{k_F}&=&\lim_{N\rightarrow\infty}l_BN\frac{\partial \Delta_{N,n}}{\partial n}\Bigr|_{n=1}=\alpha\beta l_B^{1-\beta}\nonumber\\
&=&l_B\lim_{N\rightarrow\infty}N\Delta_N=\frac{1}{2}l_B\beta\tilde \epsilon
\end{eqnarray}
where $\Delta_N$ is the smallest charge gap of the Jain state at $\nu=N/\left(2N+1\right)$ with $N\left(N+1\right)$ electrons, when an additional $\text{CF}_2$ is added to the $\left(N+1\right)^{\text{th}}$ CF level.
Again the gapless nature of the CFL requires $\lim_{N\rightarrow\infty}\Delta_N=0$. The two readily measurable physical quantities here are $k_F$ and $v_F$. One should note $k_F$ is independent of the microscopic details, or the specific form of $\Delta_N$. It can be measured with the quantum oscillation or the geometric oscillation of the magnetoresistance\cite{PhysRevLett.114.236406,PhysRevB.100.041112}. 

The fermi velocity $v_F$, on the other hand, depends on the microscopic details. In particular, if $\Delta_N$ decays slower than $1/N$, the fermi velocity will diverge. If $\Delta_N$ decays faster than $1/N$, then $v_F$ will be zero at the fermi surface, giving a flat band of $\text{CF}_2$. The dispersion exponent $\beta$ can be extracted if $\Delta_N$ decays exactly at $1/N$. These different types of CFL are determined entirely by the microscopic interaction between electrons, and can be extensively analysed using numerical computations. 

It is also important to emphasise we can only talk about $\text{CF}_2$ at filling factor $\nu\le 1/2$. This is obvious from our construction, and also natural given each $\text{CF}_2$ contains one electron and two fluxes, so the filling factor can never go beyond $\nu=1/2$. For filling factor $\nu\ge 1/2$, only $\text{PH-CF}_2$ are well-defined particles, which can form its own CFL which is the PH conjugate of the CFL from electrons. This is clearly demonstrated in geometric resonance measurement, and was previously explained from the numerical computation of the conventional CFL using the LLL projection. Here we show the asymmetry in the geometric resonance near $\nu=1/2$ is a direct consequence of the nature of the CFs, without the need of numerical computation or finite size scaling (see Fig.(\ref{fig5})).

\subsection{The effective Hamiltonian of the CFL}

The most well-known effective theories of the CFL are the Halperin-Lee-Read theory of massive composite fermions\cite{PhysRevB.47.7312}, and more recently the Dirac composite fermion theory by Son\cite{PhysRevX.5.031027,Son2013,Gromov2017}. In both cases, the starting point is to attach magnetic fluxes to electrons, thus the theories are manifestly \emph{not} within the LLL. It is thus useful to start with the microscopic Hamiltonian and understand how the effective theories of the CFL could emerge entirely within a single LL. The full Hamiltonian of the interacting electrons in a quantum Hall system is given as follows:
\begin{eqnarray}\label{ham}
\hat H_{\text{electron}}=\hat K\left(\hat p_i,\mathcal A_i,m\right)+\int d^2\bm q V_{\bm q}\hat\rho_{\bm q}\hat\rho_{-\bm q}
\end{eqnarray}
which is a more explicit representation of Eq.(\ref{full}). The first term $\hat K$ is the kinetic energy that defines the LLs, which depends on the particle momentum $\hat p_i$, the electromagnetic vector potential $\mathcal A_i$, and the electron mass $m$. The second term of Eq.(\ref{ham}) gives the electron-electron interaction, where $V_{\bm q}$ is the interaction in the momentum space, and $\hat\rho_{\bm q}=\sum_ie^{i\bm q\bm r_i}$ is the bare electron density operator. We work in the limit of the strong magnetic field, formally taking the cyclotron frequency $\omega_c=eB/m\rightarrow\infty$.

It is important to note that in this limit, the physical Hilbert space is a truncated space of a single LL, in which the two spatial coordinates no longer commutes. Thus the electrons are no longer point particles in this space, but ``quasiparticles" occupying a finite area, with the same quantum numbers (e.g. charge and spin) of an electron. It is these ``quasiparticles" that are the fundamental degrees of freedom in a single LL. In this microscopic picture, the composite fermions are not from the flux attachment to electrons as point particles, but flux attachment to these ``quasiparticles". These $\text{CF}_q$ are fundamental degrees of freedom within the subspace of a single LL. For example, $\text{CF}_2$ are the fundamental ``fermions" in the null space of $\hat V_1$. The relationship between ``quasiparticles" and $\text{CF}_2$ is in complete analogy to the relationship between electrons and ``quasiparticles". Thus the $\text{CF}_q$ defined in this work are fundamentally different from the composite fermions in the HLR theory or the Dirac fermion theory in a subtle way.

One should also note that only $\hat K$ depends on the external electromagnetic field, thus the Hall conductivity as a response to $\mathcal A_i$ is entirely determined by the kinetic energy. For translationally invariant systems, Lorentz invariance also dictates that the Hall conductivity is given by the electron density, independent of the nature of the electron-electron interaction\cite{stevegirvinlecture}. The importance of the interaction term is physically due to its determination of the energy spectrum. If a gap arises from the interaction, then the Hall conductivity plateau can develop due to the Anderson localization in the presence of disorder. Similarly, the longitudinal resistance also depends on the energy spectrum, which will be suppressed if the temperature is smaller than the ground state gap of the energy spectrum. For example, the CFL is gapless, but away from the half filling and in the presence of a periodic potential, a small gap will open when the CF fermi wave vector is commensurate with the potential periodicity, leading to the geometric resonance observed in the experiments.

Thus all effective theories of the FQH (including the CFL) should be derived only from the interaction part of Eq.(\ref{ham}), with the understanding that the Hilbert space is a single LL. In principle, such effective theories capture the energy spectrum of the system (e.g. the dispersion of the $\text{CF}_q$), but they do not ``predict" the Hall conductivity of the system; it has to be put in by hand and fixed by the electron density of the system. Other topological indices, including the Hall viscosity, topological spins and the quasihole degeneracy, on the other hand, should be captured by the effective theories. Assuming at $\nu=1/2$, the CFL has a well-defined dispersion relation given by Eq.(\ref{dispersion}) (i.e. $\Delta_N$ scales with $1/N$), we will thus have the following:
\begin{eqnarray}\label{hameff}
\hat H_{\text{electron}}'&=&\int d^2\bm q V_{\bm q}\bar\rho_{\bm q}\bar\rho_{-\bm q}=\hat H_{\text{CF}_2}\nonumber\\
&=&\sum_i\alpha\left(\sqrt{\hat p_{\text{CF},i,x}^2+\hat p_{\text{CF},i,y}^2}\right)^\beta+\hat V_{\text{CF}}\qquad
\end{eqnarray}
where $\bar\rho_{\bm q}$ is the guiding center density operator of electrons within a single LL satisfying the GMP algebra\cite{girvin1986magneto}:
\begin{eqnarray}\label{gmp}
[\bar\rho_{\bm q_1},\bar\rho_{\bm q_2}]=2i\sin\frac{\bm q_1\times\bm q_2l_B^2}{2}\bar\rho_{\bm q_1+\bm q_2}
\end{eqnarray}
$\hat p_{\text{CF}}$ is the momentum operator of a single $\text{CF}_2$, and $\hat V_{\text{CF}}$ gives the interaction between $\text{CF}_2$. Note that the Hilbert space of $\text{CF}_q$ is only a subspace of the single LL, and here we are assuming $\hat H_{\text{electron}}'$ or $\hat H_{\text{CF}_2}$ does not mix that subspace with states in the single LL outside of that subspace. For realistic interaction such mixing should be non-zero especially in numerical calculations, which is the source of ``slight" violation of the Luttinger theorem for finite systems\cite{PhysRevLett.115.186805}.

In general with two-body interaction between electrons, the effective interaction between $\text{CF}_2$ can be exactly derived, leading to both two-body and few-body interactions (see Sec.~\ref{fractalmodelham}). We thus have the following general expression:
\begin{eqnarray}
\hat V_{\text{CF}}=\sum_{n=2}^{\infty}\sum_{k,\alpha_k}\lambda_{k,\alpha_k,n}\tilde V_{k,\alpha_k}^{\text{n-bdy}}
\end{eqnarray}
where $\tilde V_{k,\alpha_k}^{\text{n-bdy}}$ is the n-body pseudopotential projecting into the sector of the total relative angular momentum $k$ of a cluster of $n$ $\text{CF}_2$, while $\alpha_k$ labels the degeneracy of such pseudopotentials. Note that $\hat V_{\text{CF}}$ is not the full effective interaction between $\text{CF}_2$ (instead $\hat H_{\text{CF}_2}$ is the full effective interaction). Rather $\hat V_{\text{CF}}=\hat H_{\text{CF}}-\hat H_{\text{CFL}}$, where $\hat H_{\text{CFL}}$ is the model Hamiltonian of the Jain state at $\nu=N/\left(2N+1\right)$ in the limit of $N\rightarrow\infty$.

If we increase the magnetic field from $\nu=1/2$, this is equivalent to the formation of the Jain state from a vacuum with a finite number of magnetic fluxes (in contrast to zero magnetic flux at exactly $\nu=1/2$). We can thus naturally treat these additional magnetic fluxes as a gauge field coupling to the $\text{CF}_2$, with the minimal coupling leading to $\hat p_{\text{CF,i,a}}\rightarrow \hat p_{\text{CF,i,a}}-|e|\tilde{\mathcal A}_a$ with the resulting LL-like energy spectrum which we dub as the CF levels. Note that the effective coupling constant $|e|$ is distinct from the electron charge, since $\tilde{\mathcal A}$ is an effective gauge field distinct from the electromagnetic gauge. There is no response of $\hat H'_{\text{electron}}$ or $\hat H_{\text{CF}_2}$ in Eq.(\ref{hameff}) to the electromagnetic vector potential $\mathcal A$. The only information of the background magnetic field comes from the magnetic length $l_B$ appearing in the GMP algebra of Eq.(\ref{gmp}).

\section{The FQH fractals}\label{fractal}

A rigorous mathematical construction of the $\text{CF}_q$ as fermionic particles forming an orthonormal basis of Slater determinants allows us to understand many FQH states in a unified manner. Let us take the CFL phase at $\nu=1/4$ as an example, which is experimentally accessible. In the usual CF description, it is a CFL of $\text{CF}_4$. There is no apparent PH symmetry and the numerical exposition of this state is technically more demanding. However, now we can easily understand the $\nu=1/4$ phase as the CFL of $\text{CF}_2$ at filling factor $\nu^*=1/2$. Formally, Eq.(\ref{cfroot}) describing the electron filling at $\nu=1/2$ is a linear combination of electronic monomials or Slater determinant states. In each of the Slater determinant, we can reinterpret electrons as $\text{CF}_2$, leading to the same state of Eq.(\ref{cfroot}) as a linear combination of the Slater determinant of $\text{CF}_2$ (where each digit ``1" indicates occupation of a $\text{CF}_2$ instead of an electron). This state describes the CFL at filling factor $\nu=1/4$, which has the identical physics of the CFL at $\nu=1/2$, just with a different type of fermions.

Since every CF Slater determinant is a well-defined linear combination of the electron wavefunctions, the wavefunction of the CFL at $\nu=1/4$ can be unambiguously constructed as a state automatically within a single LL, and one can show it has high overlap with the exact ground state of the LLL Coulomb interaction. In fact, if we have a model Hamiltonian for electron-electron interaction of the CFL at $\nu=1/2$ (e.g. $\hat V_{\text{LLL}}$), it is also the model Hamiltonian of the CFL at $\nu=1/4$, if it describes the effective interaction between $\text{CF}_2$. We can also unambiguously map this CF Hamiltonian to an electron Hamiltonians that give the exact physics (including the non-universal dynamics) of the $\nu=1/4$ CFL corresponding to the CFL at $\nu=1/2$, which we will give more details in Sec.~\ref{fractalmodelham}.

Similarly, the Laughlin $\nu=1/5$ state can be identified as the Laughlin state of $\text{CF}_2$ at $\nu^*=1/3$. It is easy to perform the PH conjugate of $\text{CF}_2$ , leading to the anti-Laughlin state of CFs at $\nu^*=2/3$, which corresponds to the FQH state of electrons at $\nu=2/7$. This PH conjugation is well-defined in the null space of $\hat V^{\text{2bdy}}_1$. Within this null space, the $\nu=1/4$ CFL state is also PH symmetric to a high level of accuracy, completely analogous to the CFL state at $\nu=1/2$ within the full Hilbert space of a single LL. The PH conjugation of $\text{CF}_q$ is very useful in understanding the chirality of different graviton modes in FQH systems\cite{multiplegraviton}.

We will now formally put electrons and composite fermions on equal footing, using $\text{CF}_q$ to denote a composite fermion from the binding of one electron with $q$ (even) fluxes, and the electrons in a single LL are the special case of $\text{CF}_0$. All these fermions are now well-defined microscopic objects related to each other by a unitary transformation. Without loss of generality we focus on the LLL. Microscopically, undressed $\text{CF}_q$ only exists within the null space of Eq.(\ref{hq}) or eigenstates of Hamiltonians of the form Eq.(\ref{hh}), and are non-interacting fermions with such a Hamiltonian. On the sphere, each $\text{CF}_q$ is a spinor of $S=q/2+n_\Lambda$ in the vacuum, with only one effective magnetic flux from the coupling of the cyclotron angular momentum (associated to different LLs) to the curvature of the sphere\cite{PhysRevLett.69.953}.  Here $n_\Lambda\ge 0$ is the CF level index (or the LL index for $q=0$), and this corresponds to the topological ``cyclotron shift" of $s_{\text{cf}_q}=2S+1$. For $q=0$ the cyclotron shift\cite{PhysRevLett.69.953} of electrons in the LLL is $\bar s_{\text{cf}_0}=1$, which is related to the LL index and is generally irrelevant for the FQHE. For $q>0$ the topological shift of $\bar s_{\text{cf}_q,n_\Lambda}=q+1+2n_\Lambda$ is analogous to the cyclotron shift of the electrons, but the part of $s_{\text{cf}_q,n_\Lambda}=\bar s_{\text{cf}_q,n_\Lambda}-\bar s_{\text{cf}_0}=q+2n_\Lambda$ is no longer associated with different LLs, so they are relevant to the guiding center topological shift of the FQHE\cite{PhysRevB.84.085316}. As a general description.  we start with an arbitrary QH state $|\psi\rangle_q$ in the $\text{CF}_q$ basis at the filling factor $\nu_{t}$ of the top CF level (so $0<\nu_{\text{cf}_q}\le 1$) and topological shift $s_{t}$ (including the integer QH case with $\nu_{t}=1$ and thus $s_{t}=0$), so we have the following relationship:
\begin{eqnarray}\label{cfphase}
N_o^{\text{cf}_q}=\nu_{t}^{-1}N_{\text{cf}_q}-s_{t}
\end{eqnarray}
Here $N_{\text{cf}_q}$ is the number of $\text{CF}_q$ in the top CF level, and $N_o^{\text{cf}_q}$ the number of single $\text{CF}$ orbitals in the top CF level, with $N_\phi^{\text{cf}_q}=N_o^{\text{cf}_q}-s_{\text{cf}_q,n_\Lambda}$ be the number of effective magnetic fluxes felt by the $\text{CF}_q$, where $n_\Lambda$ is the top CF level index. The filled CF levels thus contain $n_\Lambda\left(N_\phi^{\text{cf}_q}+n_\Lambda+q-1\right)$ fermions. The total number of $\text{CF}_q$, the same as the electron number, is given by $N_{\text{cf}_0}=N_{\text{cf}_q}+n_\Lambda\left(N_\phi^{\text{cf}_q}+n_\Lambda+q-1\right)$, and thus there are $N_\phi^{\text{cf}_0}=N_\phi^{\text{cf}_q}+qN_{\text{cf}_0}$ effective magnetic fluxes felt by $\text{CF}_0$, as each $\text{CF}_q$ contains $q$ fluxes. This leads to the following relation:
\begin{eqnarray}\label{cftoe}
N_o^{\text{cf}_0}&&=N_\phi^{\text{cf}_0}=\left(\frac{1}{n_\Lambda+\nu_{t}}+q\right)N_{\text{cf}_0}+\nonumber\\
&&\frac{n_\Lambda\left(n_\Lambda+s_t-1\right)}{n_\Lambda+\nu_{t}}-s_t-s_{\text{cf}_q,n_\Lambda}\quad\quad
\end{eqnarray}
Here $N_\phi^{\text{cf}_0}$ is the number of single particle orbitals for $N_{\text{cf}_0}$ (i.e. electrons in the LLL), since $s_{\text{cf}_0,0}=0$; however do note the physical number of magnetic fluxes is $N_\phi=N_\phi^{\text{cf}_0}-\bar s_{\text{cf}_0}$, to account for the electron cyclotron shift. Thus Eq.(\ref{cftoe}) gives the following electron filling and topological shift:
\begin{eqnarray}
&&\nu=\left(\frac{1}{n_\Lambda+\nu_{t}}+q\right)^{-1}\label{index1}\\
&&s=s_t+s_{\text{cf}_q,n_\Lambda}-\frac{n_\Lambda\left(n_\Lambda+s_t+1\right)}{n_\Lambda+\nu_{t}}\label{index2} \quad
\end{eqnarray}
for the FQH phase of Eq.(\ref{cfphase}).

We emphasise that the topological phase of Eq.(\ref{cfphase}) is an IQH of $\text{CF}_q$ for $\nu_{t}=1$ (the top CF level completely filled, with all lower CF levels completely filled), and a FQH of $\text{CF}_q$ when $\nu_{t}<1$. They are all FQH phase of electrons unless $q=0,\nu_{t}=1$. The ground states and quasihole states can be explicitly expressed as linear combination of electron (or $\text{CF}_0$) monomials following the procedures in Eq.(\ref{unitary}). Let such a typical state be $|\psi\rangle_{\text{cf}_q}$, which we can express as:
\begin{eqnarray}\label{linearcombi}
|\psi\rangle_{\text{cf}_q}&=&\sum_\lambda c_\lambda|m_\lambda\rangle_{\text{cf}_q}\nonumber\\
&=&\sum_\lambda c_\lambda d^{\text{cf}_q}_{\lambda,\lambda'}|m_{\lambda'}\rangle_{\text{cf}_0}
\end{eqnarray}
where $|m_\lambda\rangle_{\text{cf}_q}$ are the monomials of $\text{CF}_q$ (note for $\nu_{t}=1$, $|\psi\rangle_{\text{cf}_q}$ itself is a monomial). For well-known FQH states (here for $\text{CF}_q$) such as the Laughlin or the Moore-Read states, etc, $c_\lambda$ are well-known and can be readily computed. On the other hand, $d^{\text{cf}_q}_{\lambda,\lambda'}$ can be computed from Eq.(\ref{unitary}). 

The key message here is for any pair of $\left(\nu_t,s_t\right)$ that characterises a known topological phase of electrons (Abelian or non-Abelian), we can apply it to Eq.(\ref{cftoe}) with any even non-negative integer $q$ (including $q=0$, which is for electrons), and non-negative integer $n_\Lambda$. This gives the same topological phase of $\text{CF}_q$. Each pair of $\left(q,n_\Lambda\right)$ corresponds to a particular FQH phase at electron filling and shift given by Eq.(\ref{index1},\ref{index2}). All ground states and the quasihole states of such an FQH phase have well-defined model wavefunctions given by Eq.(\ref{linearcombi}), and they also have exact model Hamiltonians derived from the known model Hamiltonians for the case of electrons, or $\text{CF}_0$.

For fermions we can always take particle-hole (PH) conjugation, as discussed in Sec.\ref{jain}. When we take the PH conjugation of $\text{CF}_q$ within the $n_\Lambda^{\text{th}}$ CF level, we obtain the new topological phase by taking $\nu_t\rightarrow 1-\nu_t,s_t\rightarrow -s_t$ in Eq.(\ref{cfphase}) and correspondingly the electron topological indices in Eq.(\ref{index1},\ref{index2}). Different values of $\left(n_\Lambda,q,\nu_t,s_t\right)$ can describe the identical topological phase. An example of this is that the Laughlin phase at $\nu_t=1/3$ of $\text{CF}_2$ (i.e. $\left(n_\Lambda,q,\nu_t,s_t\right)=(0,2,1/3,2)$) is the same as the IQH phase of $\text{CF}_4$ at $\nu_t=1$ (i.e. $\left(n_\Lambda,q,\nu_t,s_t\right)=(0,4,1,0)$), which is the Laughlin $\nu_t=1/5$ phase of $\text{CF}_0$ (i.e. electrons, or $\left(n_\Lambda,q,\nu_t,s_t\right)=(0,0,1/5,4)$). 

\begin{figure}[ht]
\begin{center}
\includegraphics[width=8cm]{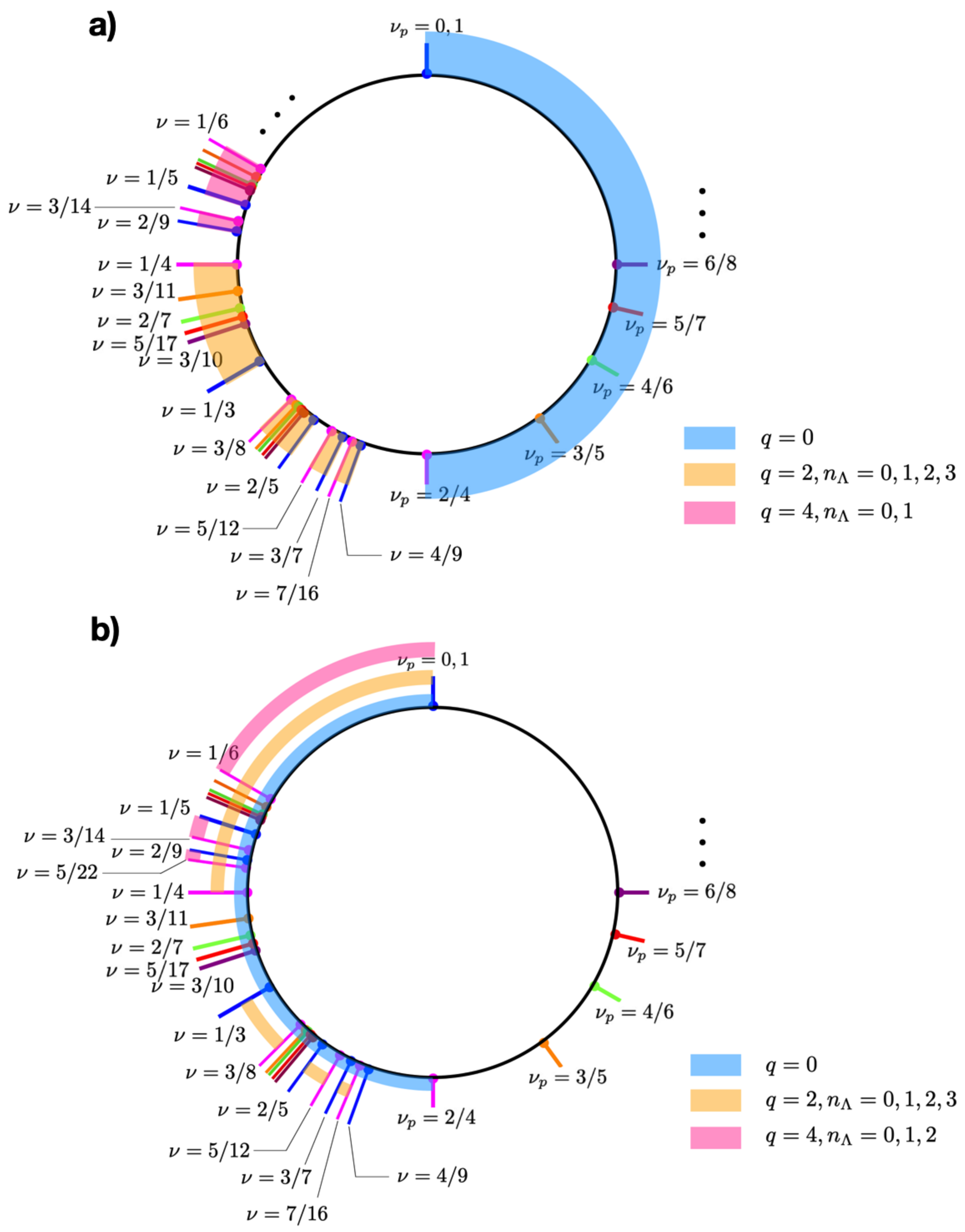}
\caption{a). The principle filling factors of the Read-Rezayi series (given by $\nu_p$ and only a few are labeled) occupies the right semicircle. Together with all other possible FQH states at the right semicircle (not labeled), they can be mapped to different intervals on the left semicircle; b). The entire left semicircle can also be mapped to different intervals within the left semicircle, clearly demonstrating a fractal structure for the distribution of the states. Due to space constraint, only a few FQH states of the electron filling factor are labeled in the figure.}
\label{fig7}
\end{center}
\end{figure}

From this perspective, all Laughlin states and the associated Jain series of $\nu=\left(n_\Lambda+1\right)/\left(1+q\left(n_\Lambda+1\right)\right)$ are ``physically equivalent" to the IQHE with $\nu_t=1$, as they can be obtained from Eq.(\ref{index1}) with different values of $n_\Lambda$ and $q$. These are all IQH states of different types of $\text{CF}_q$ (their PH conjugates within a single LL are IQH of different types of $\text{PH-CF}_q$), which are all Abelian FQH states and we can use $\nu_t=1$ to represent all of them. Note the electron filling factor for all of them are smaller than $1/2$. It is thus no coincidence that the next phase in the Read-Rezayi series\cite{read1999beyond} is the Moore-Read state at $\nu=1/2$, which is non-Abelian and fundamentally different from the $\nu_t=1$ family. We thus propose the principle FQHE series to be those from the Read-Rezayi series with $\nu_p=n_r/\left(n_r+2\right)$ with $n_r=2,3, ...$ and $s_t=2$. Here $n_r=2$ corresponds to the Moore-Read state, and $n_r=3$ corresponds to the Fibonacci state\cite{PhysRevB.94.075108,PhysRevB.95.115136}, etc. As we argued here, the $n_r=1$ case gives the Laughlin state that is equivalent to the case of $\nu_p=1$. The secondary series of FQH states comes from Eq.(\ref{index1},\ref{index2}) with the principle $\nu_p$ and different values of $n_\Lambda,q$, as shown in Fig.(\ref{fig7}).

It is interesting to see that the distribution of the secondary filling factors has fractal characteristics at $\nu_t\le 1/2$. Actually as shown in Fig.(\ref{fig7}a), all FQH states with $1/2\le\nu\le 1$ (including the Read-Rezayi series with $n_r>1$, the PH conjugate of states with $\nu<1/2$, as well as other possibly unknown FQH states) can be mapped to different intervals on the left semicircle, each interval obtained by different values of $q,n_\Lambda$. The $\nu_p=1$ state is mapped to the familiar Jain series, and all of them are Abelian. The $\nu_p=2/4$ Pfaffian state is mapped to Pfaffians of $\text{CF}_q$ at different CF levels with $\nu=\left(2n_\Lambda+1\right)/\left(2+2n_\Lambda q+q\right)$, all of them are non-Abelian states just like the well-known Moore-Read state. The same is true for other non-Abelian Read-Rezayi states, such as the Fibonacci state at $\nu_p=3/5$. There are also other Abelian states in each of the interval. For example the PH conjugate of the Jain state at $\nu=3/7$ is another Abelian state at $\nu=4/7$. The corresponding $\text{CF}_q$ states at $\nu=\left(7n_\Lambda+4\right)/\left(7+7n_\Lambda q+4q\right)$ are all Abelian and correspond to the same family of quantum Hall states as the $\nu_p=1$ IQH state. 

For the entire semicircle on the left with $0<\nu<1/2$, all these states at electron filling can also be mapped to the $\text{CF}_q$ states at the same filling. Using Eq.(\ref{index1},\ref{index2}), again the entire left semicircle can be mapped ``verbatim" to different intervals, some of them illustrated in Fig.(\ref{fig7}b). If we zoom into each interval, we will see replica of the left semicircle in terms of the distribution of the FQH states. Such an intricate hierarchical and fractal-like distribution of strongly correlated topological order is worth further investigation, and it is important to note that all states discussed here can be classified into families. States in each family are physically equivalent to one of the Read-Rezayi states, thus fundamentally speaking equivalent to one another, even though they can have different conventional topological indices (i.e. the electron filling factor and topological shift).  

\subsection{The Model Hamiltonians}\label{fractalmodelham}

With Coulomb based interaction in simple experimental systems, most of the FQH states constructed above will not be realised. However, all of them are still well-defined topological phases that can in principle be realised, because we can construct model Hamiltonians for each one of them, with dynamical properties (e.g. the incompressibility gap) identical to those of the principle filling factors, or the Read-Rezayi series at $\nu_p$. Such model Hamiltonians can be exactly constructed in principle, because all FQH states given in Fig.(\ref{fig7}) are related to the many-body electron wavefunctions of the well-known Read-Rezayi series via a unitary transformation. 

Let the model Hamiltonian describing the interaction between $\text{CF}_q$ of a particular FQH phase given by Eq.(\ref{cfphase}) be
\begin{eqnarray}\label{cfefham}
\hat H_{\text{CF}_q}=\sum_{k',n',\alpha'}\lambda_{k',n',\alpha'}\tilde V_{k',\alpha'}^{n'\text{-bdy}}
\end{eqnarray}
It is sufficient to illustrate in details the process using a Read-Rezayi state of $\text{CF}_q$, with $\nu_t=n_r/\left(n_r+2\right), s_t=2$. The model Hamiltonian between the $\text{CF}_q$ (not the electrons) is the $\left(n_r+1\right)-$body pseudopotential $\hat H_{\text{CF}_q}=\tilde V_{k_{n_r},\beta}^{\left(n_r+1\right)-\text{bdy}}$ with $k_{n_r}=n_r\left(n_r+1\right)/2$ and $\beta$ labels the degeneracy of such pseudopotentials. This Hamiltonian projects into a cluster of $n_r+1$ $\text{CF}_q$ with total relative angular momentum $k_{n_r}$ between $\text{CF}_q$ (not electrons). With $k_{n_r}=n_r\left(n_r+1\right)/2$ the pseudopotential is unique so we have $\beta=1$, but we will still leave $\beta$ in the notation. Our goal is to derive the effective interaction between electrons that gives the identical energy spectrum as that of the Read-Rezayi state from $\tilde V_{k_{n_r},\beta}^{\left(n_r+1\right)-\text{bdy}}$.

Let us first set up the notations. We use $|k,n,\alpha\rangle_{\text{CF}_q}$ to denote a set of orthonormal quantum state containing $n$ numbers of $\text{CF}_q$, with total relative angular momentum $k$, and $\alpha=1,2,\cdots \alpha_{n,k}$ labels the degeneracy of such states. We have established from the fermionization scheme that these states can be mapped to many-body electron states with the same quantum number within the conformal Hilbert space $\mathcal H_q$, which we can denote as $|k,n,\alpha\rangle_{e,\mathcal H_q}$. For $n_\Lambda=0$, $\mathcal H_q$ is the null space of Eq.(\ref{hq}); for $n_\Lambda>0$ it is the null space of the model Hamiltonians in the form of Eq.(\ref{hh}). These many-body electron states are well-defined linear combinations of electron monomials, each containing $n$ electrons, with coefficients that can be readily obtained. In particular we can have the following expression:
\begin{eqnarray}
|k,n,\alpha\rangle_{e,\mathcal H_q}=\sum_{k',\alpha'}\tilde\lambda^{k'\alpha'}_{n,k\alpha}|k',n,\alpha'\rangle_{e}
\end{eqnarray}
where $|k',n,\alpha'\rangle_{e}$ are quantum states of $n$ electrons, with total relative angular momentum between \emph{electrons} as $k'$. The coefficients of this basis transformation, $\lambda^{k'n\alpha'}_{kn\alpha}$, can be computed from Eq.(\ref{unitary}). For example, it is easy to check that for $n=2$, we have $\tilde\lambda^{k'\alpha'}_{2,k\alpha}=\delta^{\alpha'}_{\alpha}\delta^{k'}_{k+q}$.

Let the effective interaction between electrons be given in the general form as follows:
\begin{eqnarray}\label{effectiveh}
\hat H_e=\sum_{k',n',\alpha'}\lambda_{k',n',\alpha'}\hat V_{k',\alpha'}^{n'\text{-bdy}}
\end{eqnarray}
We need to determine all coefficients of $\lambda_{k',n',\alpha'}$ such that:
\begin{eqnarray}\label{hmap}
&&_{\text{CF}_q}\langle k,n,\alpha|\tilde V_{k_{n_r},\beta}^{n_r-\text{bdy}}|k,n,\alpha\rangle_{\text{CF}_q}\nonumber\\
&&=_{e,\mathcal H_q}\langle k,n,\alpha|\hat H_e|k,n,\alpha\rangle_{e,\mathcal H_q}
\end{eqnarray}
Both sides of Eq.(\ref{hmap}) can be readily computed. The LHS is obviously zero for $n<n_r+1$; for $n=n_r+1$ it equals to the vector $\vec{\Delta}_{n_r+1,\left(k,\alpha\right)}=\delta_{k,k_{n_r}}\delta_{\alpha,\beta}$, where we treat $\left(k,\alpha\right)$ as the index of the vectors (and later for the matrices). The RHS is equal to:
\begin{eqnarray}
&&_{e,\mathcal H_q}\langle k,n_r+1,\alpha|\hat H_e|k,n_r+1,\alpha\rangle_{e,\mathcal H_q}\nonumber\\
&&=\sum_{k',\alpha'}\lambda_{k',n_r+1,\alpha'}\overleftrightarrow{M}_{n_r+1,\left(k,\alpha\right)}^{\left(k',\alpha'\right)}
\end{eqnarray}
with the matrix $\overleftrightarrow{M}_{n_r+1,\left(k,\alpha\right)}^{\left(k',\alpha'\right)}=|\tilde\lambda_{n_r+1,k\alpha}^{k'\alpha'}|^2$. Defining a vector $\vec \lambda_{n_r+1,\left(k',\alpha'\right)}=\lambda_{k',n_r+1,\alpha'}$ we immediately obtain $\vec\lambda_{n_r+1}=\left(\overleftrightarrow M_{n_r+1}\right)^{-1}\vec\Delta_{n_r+1}$, and thus all values of $\lambda_{k',n_r+1,\alpha'}$.

The rest of the coefficients can be obtained inductively. Assuming all $\lambda_{k',n',\alpha'}$ are known for $n'< n_0$, we now look at the case of $n=n_0$. By assumption the following quantities are known:
\begin{eqnarray}
&&C_{kn_0\alpha}=\quad_{\text{CF}_q}\langle k,n_0,\alpha|\tilde V_{k_{n_r},\beta}^{n_r-\text{bdy}}|k,n_0,\alpha\rangle_{\text{CF}_q}\label{i1}\\
&&D_{kn_0\alpha}=\quad\sum_{k',\alpha'}\sum_{n'=n_r+1}^{n_0-1}\lambda_{k',n',\alpha'}\overleftrightarrow{M}_{n',\left(k,\alpha\right)}^{\left(k',\alpha'\right)}
\end{eqnarray}
We then have:
\begin{eqnarray}\label{ii}
\vec\lambda_{n_0}=\left(\overleftrightarrow M_{n_0}\right)^{-1}\vec\Delta_{n_0}
\end{eqnarray}
with $\vec\Delta_{n_0,\left(k,\alpha\right)}=C_{kn_0\alpha}-D_{kn_0\alpha}$. In fact we can compute all the values of $\lambda_{k',n',\alpha'}$ with Eq.(\ref{i1}-\ref{ii}), by noting that $C_{k,n_r+1,\alpha}=\delta^{\alpha'}_{\alpha}\delta^{k'}_{k+q}$ and $D_{k,n_r+1,\alpha}=0$.

Thus Eq.(\ref{effectiveh}) and Eq.(\ref{cfefham}) are related by a unitary transformation, as it should be, since the electron basis and the CF basis are also related by a unitary transformation. Therefore, if we know the electron interaction Hamiltonian in the form of Eq.(\ref{effectiveh}) (e.g. the Coulomb interaction), we can then derive the effective, or residual interaction between the CFs (which is described by Eq.(\ref{cfefham})) using the same procedure.

It is important to note that Eq.(\ref{effectiveh}) describes the effective interaction within a single CF level. The full electron-electron interaction is thus given by
\begin{eqnarray}\label{fullmodel}
\hat H_{\text{model},\nu,s}=\hat H_{\text{CF}_q,n_\Lambda}+\hat H_e
\end{eqnarray}
This model Hamiltonian is actually analogous to Eq.(\ref{full}). While the first term is also a linear combination of pseudopotentials between electrons, it is very much analogous to the kinetic energy giving the Landau levels, or the first term of Eq.(\ref{full}). For $n_\Lambda=0$, $\hat H_{\text{CF}_q,n_\Lambda}=\hat H_q$ in Eq.(\ref{hq}); for $n_\Lambda>0$, $\hat H_{\text{CF}_q,n_\Lambda}$ has to be constructed in the same way as Eq.(\ref{hh}). Just like the kinetic energy giving the LLs, the first term needs to be the dominant energy scale (or sent to infinity), as the second term may introduce mixing between different CF levels, just like the interaction part of the full electron Hamiltonian.

The procedure above can be easily generalised beyond the Read-Rezayi states of $\text{CF}_q$. As long as a particular FQH phase given by Eq.(\ref{cfphase}) has a known model Hamiltonian (as a linear combination of pseudopotentials), we can derive the effective model electron-electron interaction Hamiltonian for each of the pseudopotential using the method shown above. In general, even for very simple interaction between $\text{CF}_q$ (e.g. a single pseudopotential), the corresponding model Hamiltonian between electrons consist of an infinite number of pseudopotential, though the coefficients of most of them tend to be very small. 

For the PH conjugates of the CF states within a single CF level (in contrast to within the LL), the corresponding CF model Hamiltonian can be simply obtained by taking the PH conjugate of Eq.(\ref{cfefham}). For example, if Eq.(\ref{cfefham}) is a two-body interaction, then its PH conjugate is the same two-body interaction plus a constant chemical potential which we can readily ignore. As an example, by taking $q=2$ and $\hat H_{\text{CF}_q}=\tilde V_1^{\text{2bdy}}$, the ground state is the $\nu^*=1/3$ Laughlin state of $\text{CF}_2$, or the $\nu=1/5$ Laughlin state of electrons. Its PH conjugate within a single Landau level gives the same model Hamiltonian $\hat H_{\text{CF}_q}$, where the exact ground state is the $\nu^*=2/3$ state of $\text{CF}_2$, or the $\nu=2/7$ state of electrons. Using Eq.(\ref{i1}-\ref{ii}), for $\nu=1/5$ Laughlin state we can derive the following:
\begin{eqnarray}
\hat H_{\text{CF},0}&=&\lambda\hat V_1^{\text{2bdy}}\\
\hat H_e&=&\hat V_3^{\text{2bdy}}+\delta H_e\\
\hat H_{\text{model},1/5,4}&=&\hat H_{\text{CF},0}+\hat H_e
\end{eqnarray}
Here $\delta H_e$ can be explicitly computed, but it contains an infinite number of pseudopotential terms. We can show that the null space of $\delta H_e$ is the same as the null space of $\hat V_3^{\text{2bdy}}$, and the simplest pseudopotential in $\delta H_e$ is $\hat V_9^{\text{3bdy}}$\cite{footnote1}, with the coefficient $\sim 0.78$. 

In principle we need to take $\lambda\rightarrow\infty$, but for the $\nu=1/5$ ground state and quasihole states this turns out to be not necessary since the null space of $\hat H_e$ is a proper subspace of the null space of $\hat H_{\text{CF},0}$. We can also take $\delta H_e=0$, which will not affect the ground state and the quasihole states, though that will affect the gapped excitations, which will no longer be identical to the gapped excitations of $\hat H_{\text{CF}_q}=\tilde V_1^{\text{2bdy}}$. For the exact mapping of the gapped excitations, we need to take $\lambda\rightarrow\infty$ and for $\delta H_e$ to be precisely computed and included in $\hat H_e$.

For its PH conjugate state within the CF level, $\nu=2/7$, however, $\hat H_e$ will mix the ground state and quasihole states with higher CF levels because $\nu>1/5$. This is analogous to the electron-electron interaction in FQH states mixing different LLs. The model Hamiltonian for $\nu=2/7$ thus requires $\lambda\rightarrow\infty$ (analogous to the magnetic field going to infinity in Eq.(\ref{full})) and naturally $\delta H_e$ cannot be ignored. It is interesting to notice that while $\hat H_e$ is \emph{not} PH symmetric within a single LL (due to the presence of $\delta H_e$), it is PH symmetric within a single CF (apart from a constant chemical potential), a different conformal Hilbert space. This again illustrates the qualitative similarities and quantitative differences between the Landau levels and the CF levels.  

Just like the model Hamiltonians of FQH should be equivalent in different LLs, the model Hamiltonian we derived here, i.e. $\hat H_e$, should also be equivalent in different CF levels. For example with $n_\Lambda=1$, the $\hat H_{\text{CF}_q}=\tilde V_1^{\text{2bdy}}$ Hamiltonian with $q=2$ gives the FQH state at $\nu=4/11$, and its PH conjugate within a single CF at $\nu=5/13$. In both cases, the electron-electron model Hamiltonian contains $\delta H_e$. The only term we need to change is $\hat H_{\text{CF},1}$, which is given by Eq.(\ref{hh}) that gives the non-interacting $\text{CF}_q$ in the $n_\Lambda=1$ CF level. Thus while $\hat H_{\text{model}}$ describing the electron-electron interaction always consists of $\hat H_{\text{CF}_q}$ and $\hat H_e$, the former is completely analogous to the ``kinetic energy" that dictates which CF level the dynamics of the composite fermions live in, while the latter is the ``true" interaction part defining the incompressibility gap from the CF interaction within a partially filled CF level. Numerical verification of this tends to be difficult, however, due to the small system sizes that are computationally accessible. For finite systems, at $n_\Lambda=1$ there are additional small corrections to $\delta H_e$. We expect they are not important for the topological properties of the relevant FQH states, and a more systematic study is needed to check if those small corrections vanish in the thermodynamic limit.

\section{Summary and outlook}\label{summary}

In summary, we have rigorously derived microscopic model wavefunctions and proper model Hamiltonians (within a single LL) of a large family of FQH phases, using composite fermions as well-defined quasiparticles as a generalisation of electrons. The unitary transformation between electrons and composite fermions are block diagonal in the particle number and the symmetry quantum number sectors, and the process also illustrates the physical importance of the concept of conformal Hilbert spaces. Many of the states discussed in this work have been studied in the past with the conventional CF theory, and we argue the model wavefunctions we constructed are topologically identical, with the systematic construction of model Hamiltonians that have been actively sought after, albeit unsuccessfully in the past. The entire scheme we developed also aims to unify the phenomenological CF approach with the more rigorous pseudopotential and Jack polynomial formalism, and in particular exposes subtle links between the Abelian Jain states and their non-Abelian counterparts at the same filling factor and topological shifts. 

While this work endeavours to establish an ``exact" FQH to IQH correspondence and thus to extract useful physical consequences, it is also clear, just from dimensional consideration, that an exact correspondence for any finite system is \emph{impossible}. This is because for a fixed electron number $N_e$ and orbital number $N_o$, the Hilbert space of the FQH effect in a single Landau level is finite dimensional, while that for the IQH is infinite dimensional. It is thus not possible to have a fully unitary transformation from the electron basis to any CF basis. 

Nevertheless, all the analysis in the previous sections are rigorous. In particular, all the attributes listed in Sec.~\ref{scheme} are exact statements. The important message here is, while all FQH states in the electron basis discussed in this work can be expressed as product states in the CF basis, not all product states in the CF basis has a corresponding many-body wavefunction in the electron basis. This issue has so far been ignored in this work, because we only focused on a partially filled top CF level, while all the lower CF levels are completely filled. These include only the ground states, and the quasihole/quasielectron excitations not dressed with neutral excitations. The exact FQH to IQH correspondence will no longer be valid only when \emph{neutral excitations} are present in the system. This include states containing dressed quasiholes/quasielectrons, i.e. quasiholes or quasielectrons together with neutral excitations.

The missing states for the neutral excitations, or the ``CF excitons", are well-known in the conventional CF theory\cite{PhysRevB.88.205312}, when a non-trivial CF wavefunction vanishes after the LL projection. The dimensional mismatch between the FQH and IQH Hilbert space is only explicit for finite systems, so this ``neutral excitation anomalies" may not be important in the thermodynamic limit, though this is an issue that warrants further careful studies. Neutral excitations are gapped excitations, so in most cases they will not affect the topological properties of the FQH ground states (as well as their quasiholes). However when we move away from ideal model Hamiltonians, neutral excitations may become well-structured low-lying excitations or even gapless, with rich physical consequences\cite{PhysRevLett.127.046402,PhysRevB.105.035144,multiplegraviton}. 

From a more fundamental level, our work here proposes $\text{CF}_q$ as well-defined microscopic elementary particles, each type responsible for a large family of FQH states. These include the IQH of the $\text{CF}_q$, the FQH of the $\text{CF}_q$ and their PH conjugates (both Abelian and non-Abelian), as well as the fermi liquid of the $\text{CF}_q$. We show the distribution of these FQH states emerge with a fractal structure on the real axis, with a more general classification of FQH states using the Read-Rezayi series as the principle topological phases. For example, all Abelian FQH states (at least those discussed in this work) are physically (or at least topologically) equivalent to the IQHE, even though they may have different topological indices (e.g. the filling factors). They are just IQHE in different conformal Hilbert spaces (the single LL is a special case) that are isomorphic to each other, with different types of fermions. This scheme of classification of topological phases beyond using the conventional topological indices may also be useful for the proper formulation of effective topological field theories with better support from the microscopic theories.

\begin{acknowledgments}
I thank A.C. Balram for the constructive discussions, J.K. Jain and G.J. Sreejith for the helpful comments. This work is supported by the NTU grant for Nanyang Assistant Professorship and the National Research Foundation, Singapore under the NRF fellowship award (NRF-NRFF12-2020-005), and a Nanyang Technological University start-up grant (NTU-SUG).\end{acknowledgments}

\bibliography{ref}

\end{document}